\documentclass[aps,prl,fleqn,showpacs,superscriptaddress,twocolumn,reprint]{revtex4-2}
\usepackage{textcomp}
\usepackage{bm}
\usepackage{amsmath,amsfonts,amssymb}
\usepackage{color}
\usepackage{hyperref}
\usepackage{listings}
\usepackage{float}
\usepackage{graphicx}
\usepackage{enumitem}
\usepackage[capitalise]{cleveref}
\usepackage{overpic}
\usepackage{CJKutf8}
\usepackage{verbatim}
\usepackage{epstopdf}
\usepackage{ulem}
\usepackage{makecell}

\hypersetup{
  colorlinks,
  backref,
  pdfborder=false
}

\newcommand{\IOP}{Beijing National Laboratory for Condensed Matter Physics and CAS Key Laboratory of Soft Matter Physics, Institute of Physics, Chinese Academy of Sciences, Beijing 100190, China}
\newcommand{\UCI}{Department of Physics and Astronomy, University of California, Irvine, California 92697, USA}
\newcommand{\USTC}{School of Physical Sciences, University of Science and Technology of China, Hefei, Anhui 230026, China}
\newcommand{\UCAS}{School of Physical Sciences, University of Chinese Academy of Sciences, Beijing 100049, China}
\newcommand{\SLAB}{Songshan Lake Materials Laboratory, Dongguan, Guangdong 523808, China}
\newcommand{\CEULS}{CAS Center for Excellence in Ultra-intense Laser Science, Shanghai 201800, China}

\newcommand{\ITERCN}{National MCF Energy R\&D Program}
\newcommand{\NSFC}{the National Natural Science Foundation of China}

\newcommand{\SPRP}{the Strategic Priority Research Program of Chinese Academy of Sciences}
\newcommand{\KRPFS}{the Key Research Program of Frontier Science of Chinese Academy of Sciences}

\begin{document}
\title{A New Paradigm for Fast and Repetitive Chirping of Alfv\'en Eigenmodes}

\author{Junyi \surname{Cheng}}
\affiliation{\IOP}
\affiliation{\USTC}
\affiliation{\UCAS}

\author{Wenlu \surname{Zhang}}
\email{wzhang@iphy.ac.cn}
\affiliation{\IOP}
\affiliation{\SLAB}
\affiliation{\UCAS}
\affiliation{\CEULS}
\affiliation{\USTC}

\author{Zhihong \surname{Lin}}
\affiliation{\UCI}

\author{Jian \surname{Bao}}
\affiliation{\IOP}
\affiliation{\UCAS}

\author{Chao \surname{Dong}}
\affiliation{\IOP}
\affiliation{\UCAS}

\author{Jintao \surname{Cao}}
\affiliation{\IOP}
\affiliation{\UCAS}

\author{Ding \surname{Li}}
\affiliation{\IOP}
\affiliation{\SLAB}
\affiliation{\UCAS}
\affiliation{\CEULS}

\date{\today}

\begin{abstract}

A novel 2D nonlinear dynamical paradigm is constructed to interpret the fast and repetitive frequency chirping and amplitude oscillation of Alfv\'en eigenmodes excited by energetic particles in fusion plasmas as observed in global gyrokinetic simulations.
In this non-perturbative paradigm of the collisionless phase-space dynamics, wave-particle resonant interactions cause the phase-space structure to continuously twist and fold, leading to the repetitive excitation and decay of the Alfv\'en eigenmode.
The radial
(perpendicular to the dominant wave-particle interaction)
dependence of the mode amplitude and toroidal precessional drifts of the energetic particles leads to the 2D dynamics of wave-particle interactions, which is found to be responsible for the repetitive process of formation and destruction of the mode structure.

\end{abstract}

\maketitle
\sloppy

\noindent
\\
\textbf{\large 1.} \textbf{\large Introduction} 
\\

Energetic particles,
including energetic ions and electrons produced by the fusion reaction and auxiliary heating,
can excite various Alfv\'en eigenmodes in magnetic confinement fusion plasmas such as ITER \cite{ikeda2007progress},
which may induce significant transport of energetic particles and degrade the overall plasma confinement\cite{strait1993stability,
podesta2011non,
garcia2011fast}. 
Increased energetic particle transport due to Alfv\'en eigenmodes has been correlated \cite{podesta2011non} with a fast frequency oscillation (chirping) with a sub-millisecond period (non-adiabatic regime\cite{chen2016physics}), which has been observed in many tokamak experiments \cite{pinches2004spectroscopic,gryaznevich2006perturbative,heidbrink2006weak}.
Previous theoretical studies focus on the adiabatic regime (chirping time much longer than the wave period) based on the 1D (in toroidal direction) dynamical model for the wave-particle interactions\cite{berk1996nonlinear}, which has been investigated by numerical simulations\cite{lilley2009destabilizing,pinches2004spectroscopic,candy1999nonlinear,vann2007,hezaveh2020long}.
In this 1D dynamical model\cite{berk1996nonlinear}, the relaxation/dissipation process\cite{lilley2009destabilizing} are essential to inducing the frequency chirping.

However, the repetitive fast chirping of frequency and amplitude oscillation of Alfv\'en eigenmodes
excited by energetic particles in the toroidal geometry
have been observed by both first-principle kinetic \cite{zhang2012nonlinear}
and hybrid magnetohydrodynamic\cite{wang2012nonlinear} simulations without source and sink,
which is in agreement with experiments and cannot be explained by the 1D dynamical model.
Here, we construct a novel 2D (in toroidal and radial directions) collisionless nonlinear dynamical paradigm based on the global gyrokinetic simulation result, which shows that the dynamics of fast and repetitive frequency chirping and amplitude oscillation
is a non-adiabatic\cite{zonca2015nonlinear}, non-perturbative and essentially intrinsic 2D nonlinear wave-particle resonant interactions,
where the evolution of 2D trajectories of resonant particles restores unstable distribution function.
In particular, the radial
(perpendicular to the dominant wave-particle interaction)
dependence of the mode amplitude and toroidal precessional drifts of the energetic particles leads to the 2D dynamics of wave-particle interactions,
which is found to be responsible for the repetitive fast frequency chirping.

Specifically, the current simulations using the gyrokinetic toroidal code (GTC) \cite{lin1998turbulent} find that beta-induced Alfv\'en eigenmodes (BAE) excited by energetic electrons saturate and chirp,
exhibiting various stages of nonlinear 2D dynamics due to the radial dependence of the mode amplitudes and the toroidal precessional drifts of energetic electrons.
Our paradigm of the 2D nonlinear wave-particle interactions can be applied to other Alfv\'en eigenmodes.

\begin{figure}[!ht]
\includegraphics[width=0.45\textwidth]{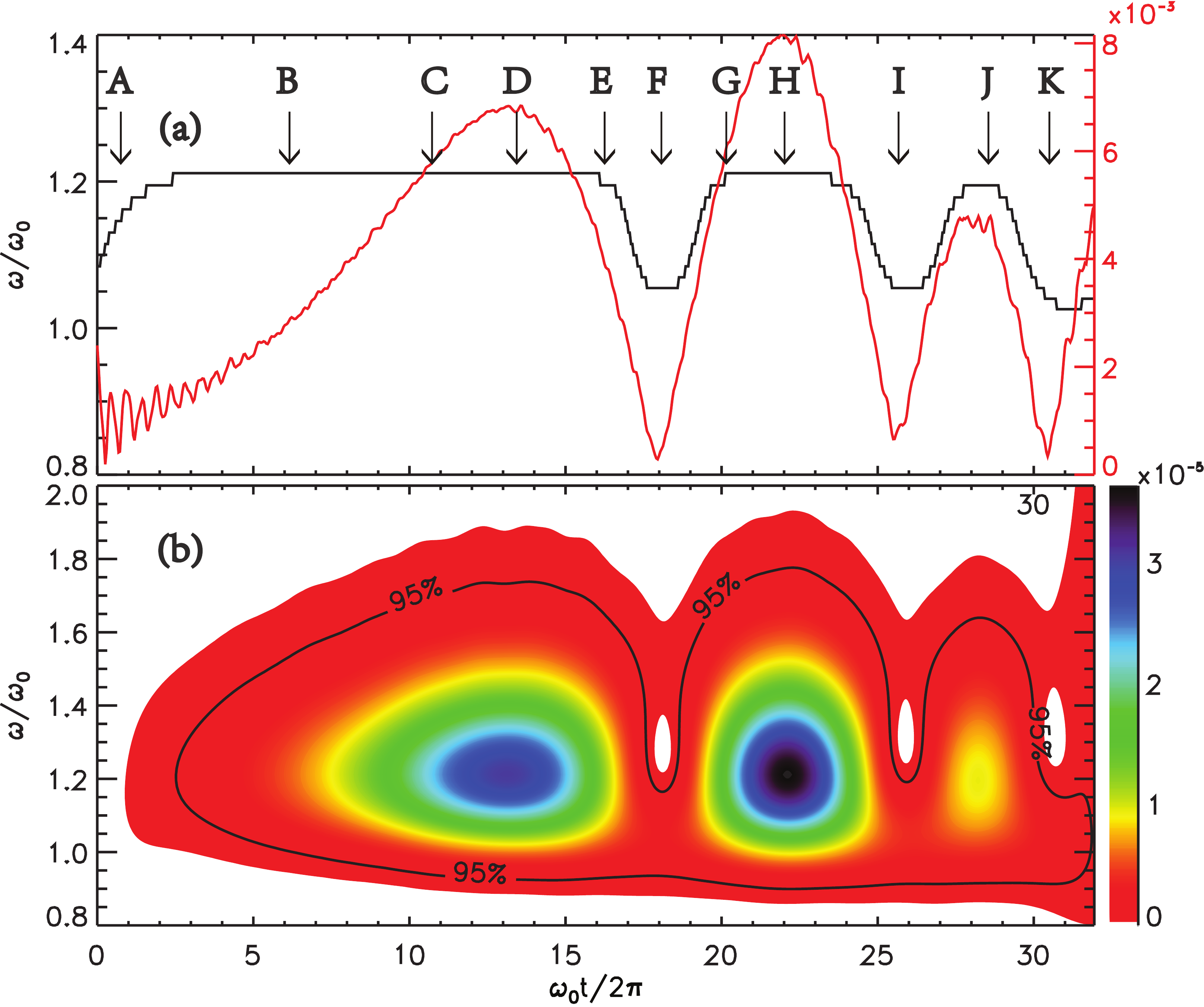}
\caption{Time evolution of electrostatic potential $|e\phi/T_e|$ (red) and the frequency $\omega$ (black) of dominant mode $(n,m)=(3,6)$ in upper panel (a) and frequency power spectrum in lower panel (b).
The color in panel (b) represents the power intensity and the $95\%$ lines indicate the significance level.
The unit of the power intensity in the panel (b) is arbitrary.
The mode stage at different times is marked with letters A-K.}\label{fig:wavelet}
\end{figure}

\begin{figure}[!ht]
  \includegraphics[width=0.15\textwidth]{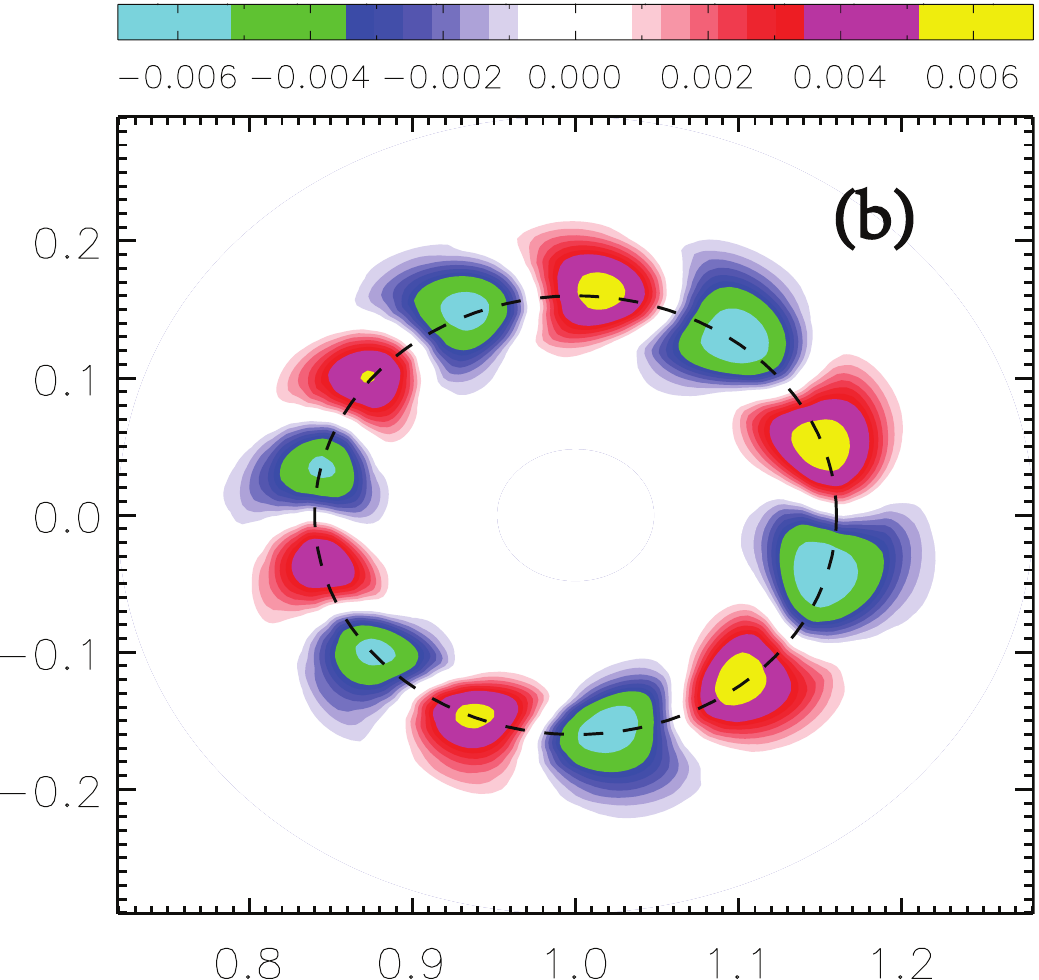}
  \includegraphics[width=0.15\textwidth]{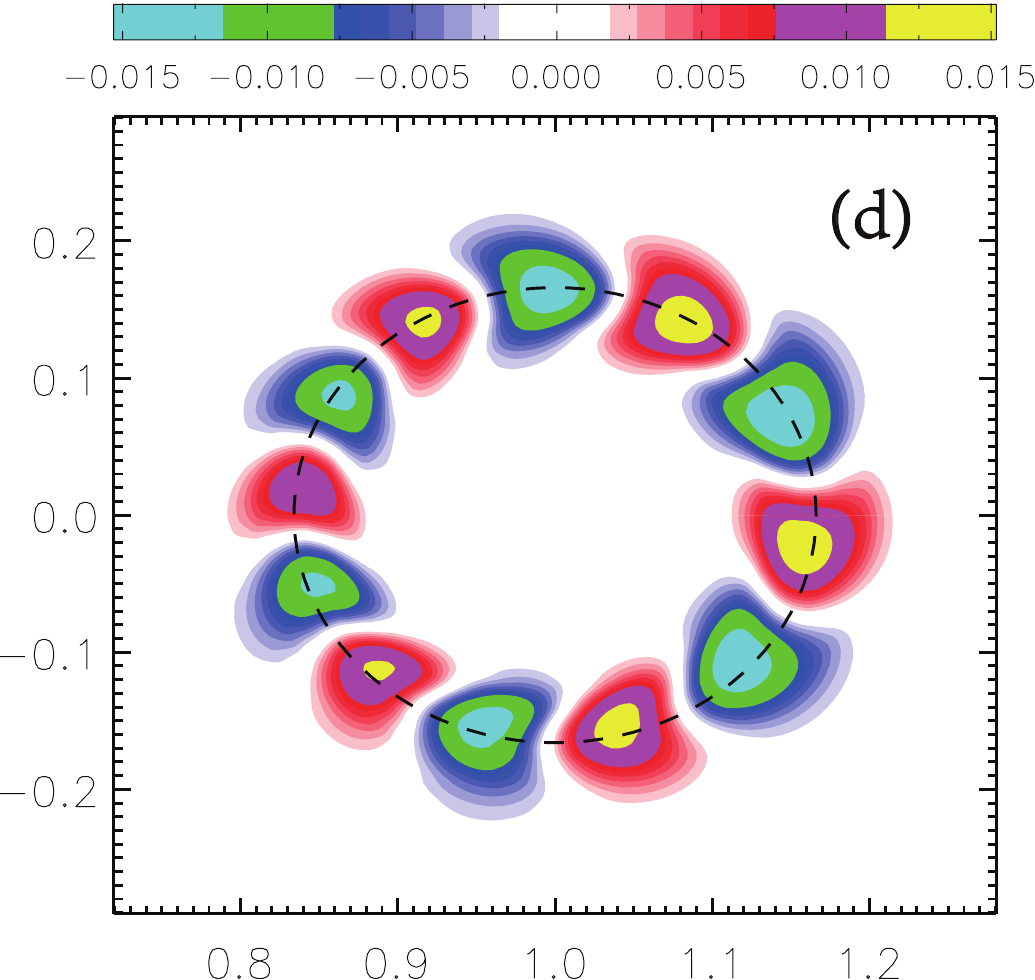}
  \includegraphics[width=0.15\textwidth]{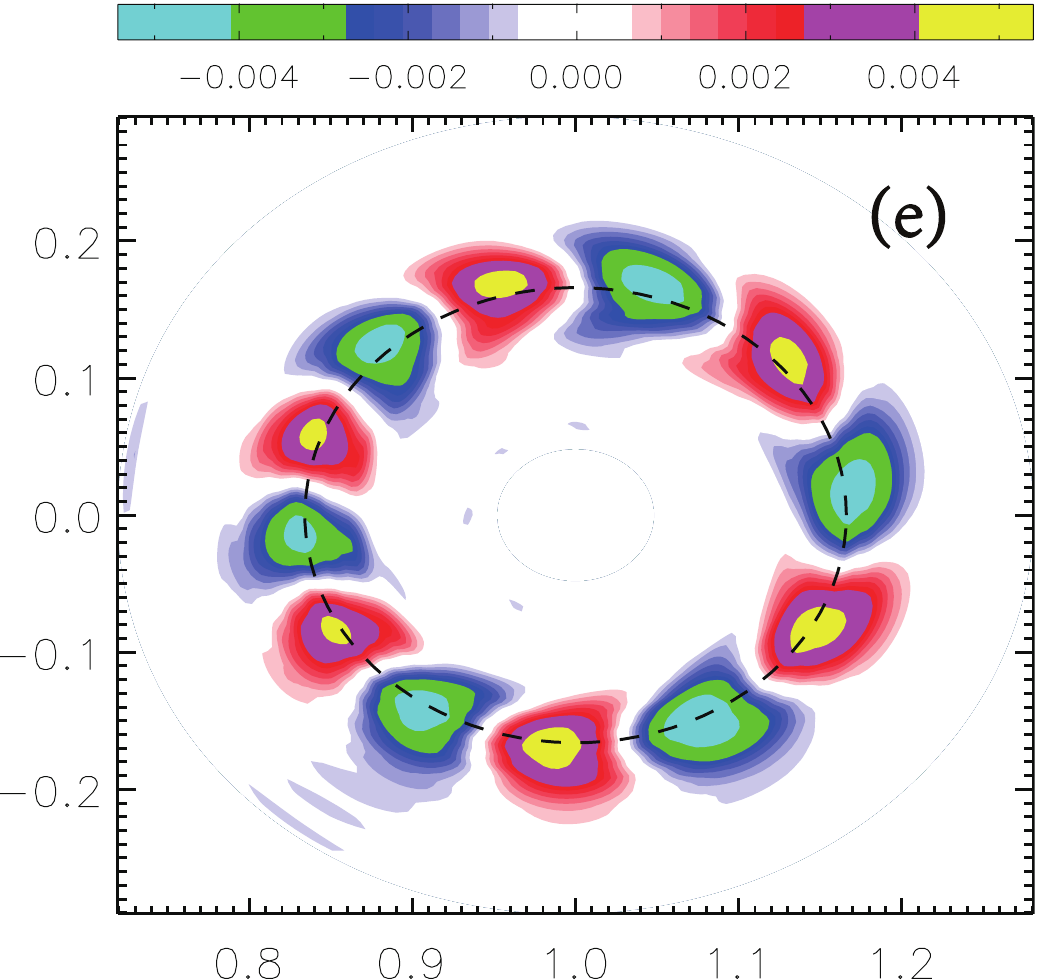}
  \includegraphics[width=0.15\textwidth]{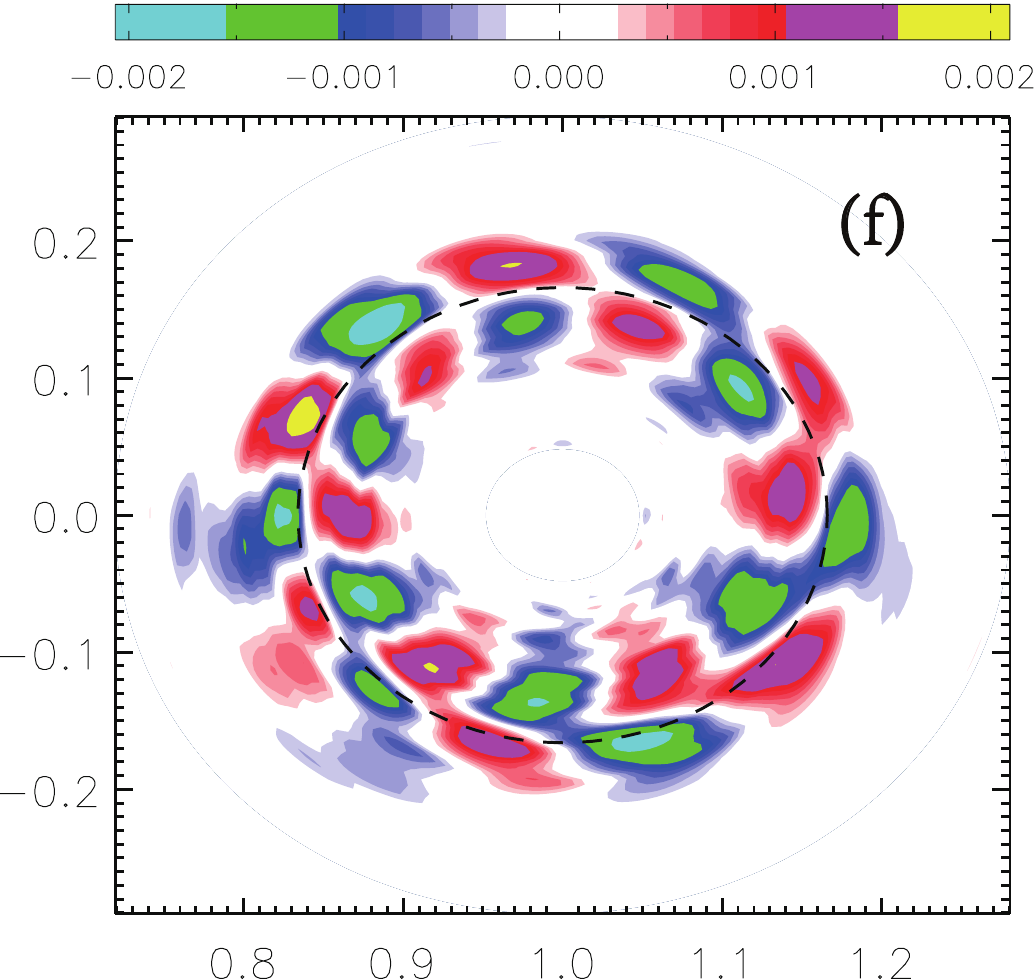}
  \includegraphics[width=0.15\textwidth]{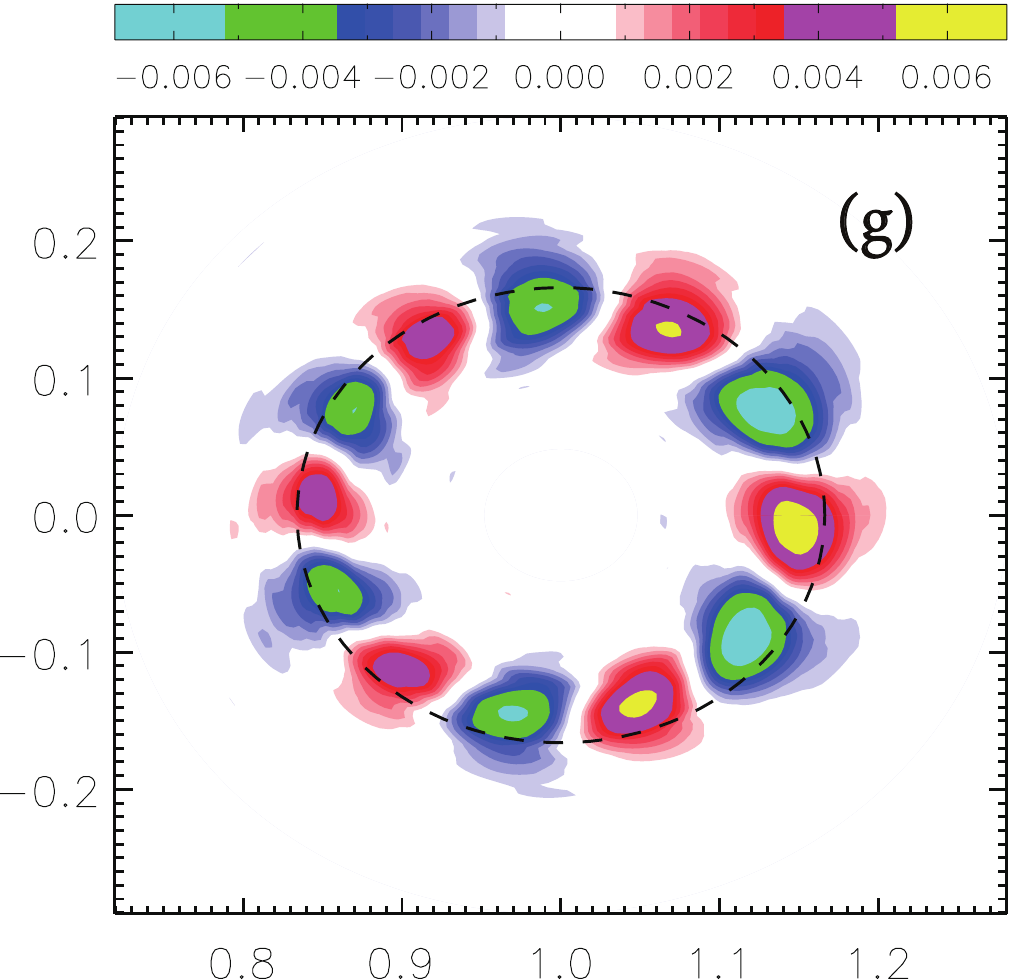}
  \includegraphics[width=0.15\textwidth]{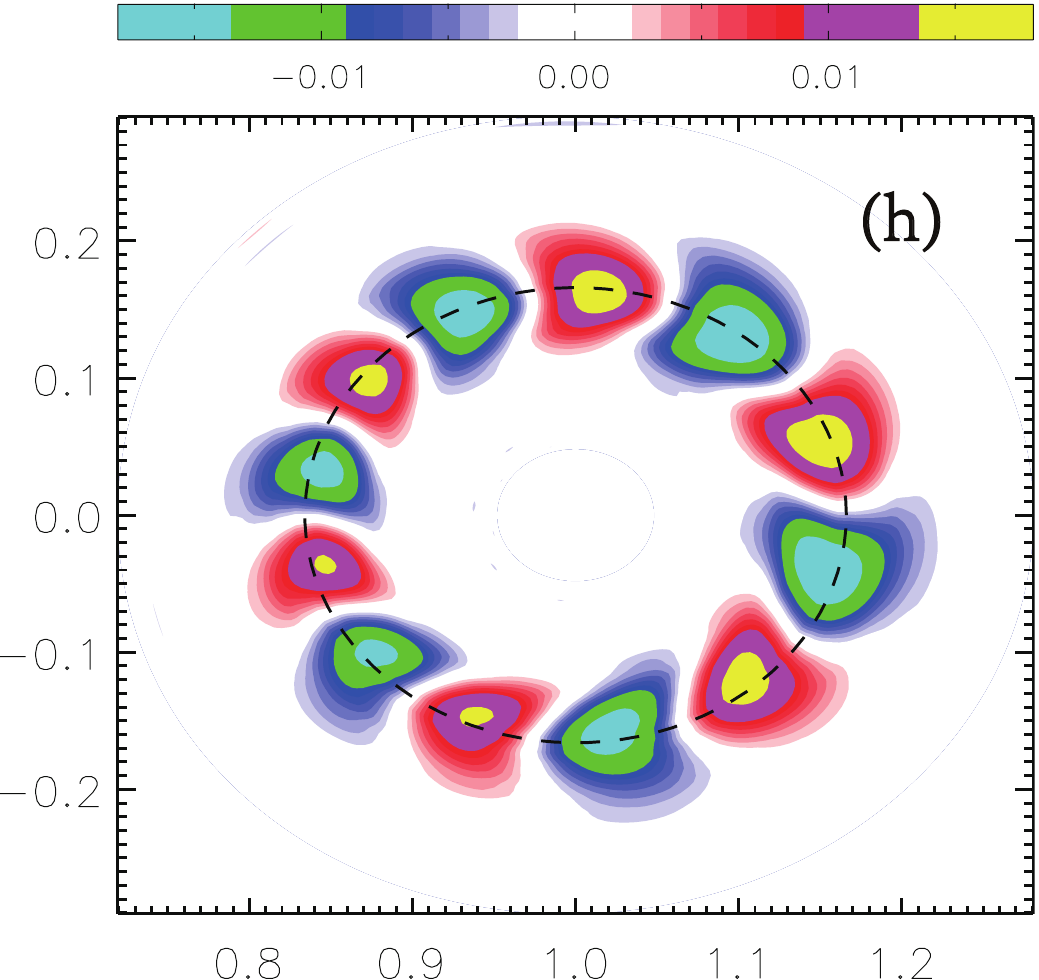}
\caption{Poloidal contour plots of electrostatic potential $e\phi/T_e$.
The dotted circle is the $q=2$ rational surface.
The $x$ axis is the major radius $R/R_0$ and the $y$ axis is the vertical distance from the midplane $Z/R_0$. 
Time steps $b-h$ correspond to $B-H$ in \cref{fig:wavelet}. }\label{fig:phi-snap}
\end{figure}

\begin{figure}[!ht]
  \includegraphics[width=0.15\textwidth]{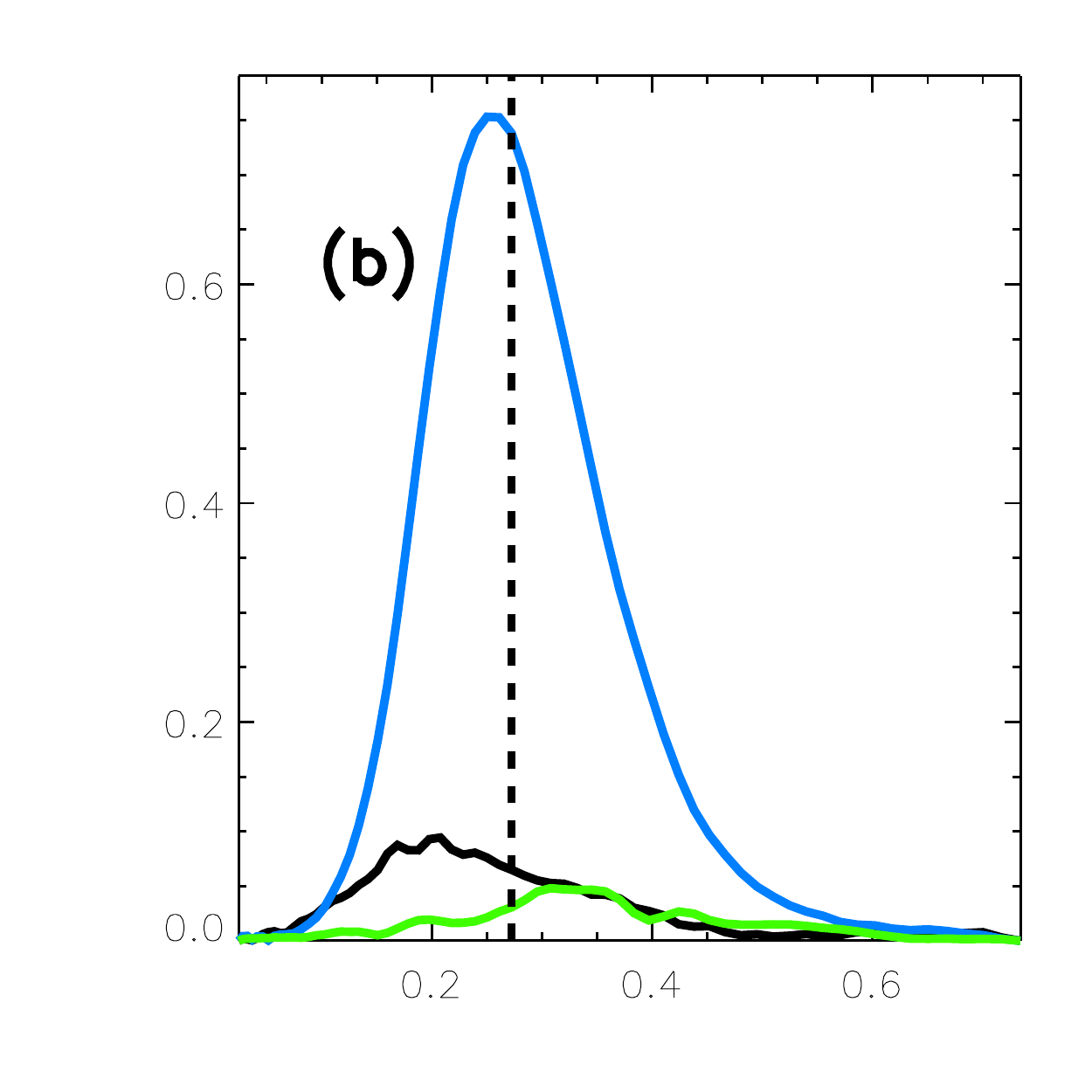}
  \includegraphics[width=0.15\textwidth]{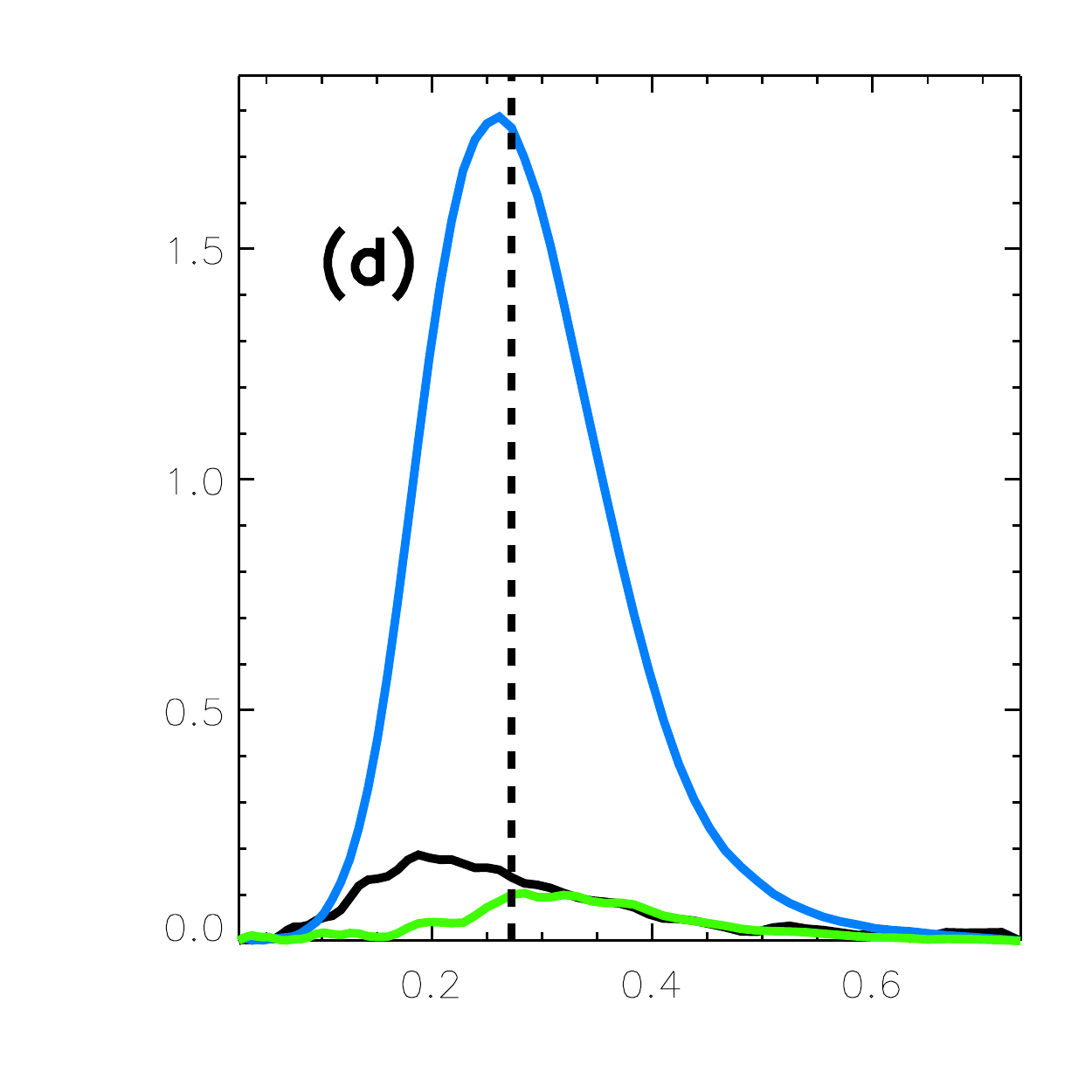}
  \includegraphics[width=0.15\textwidth]{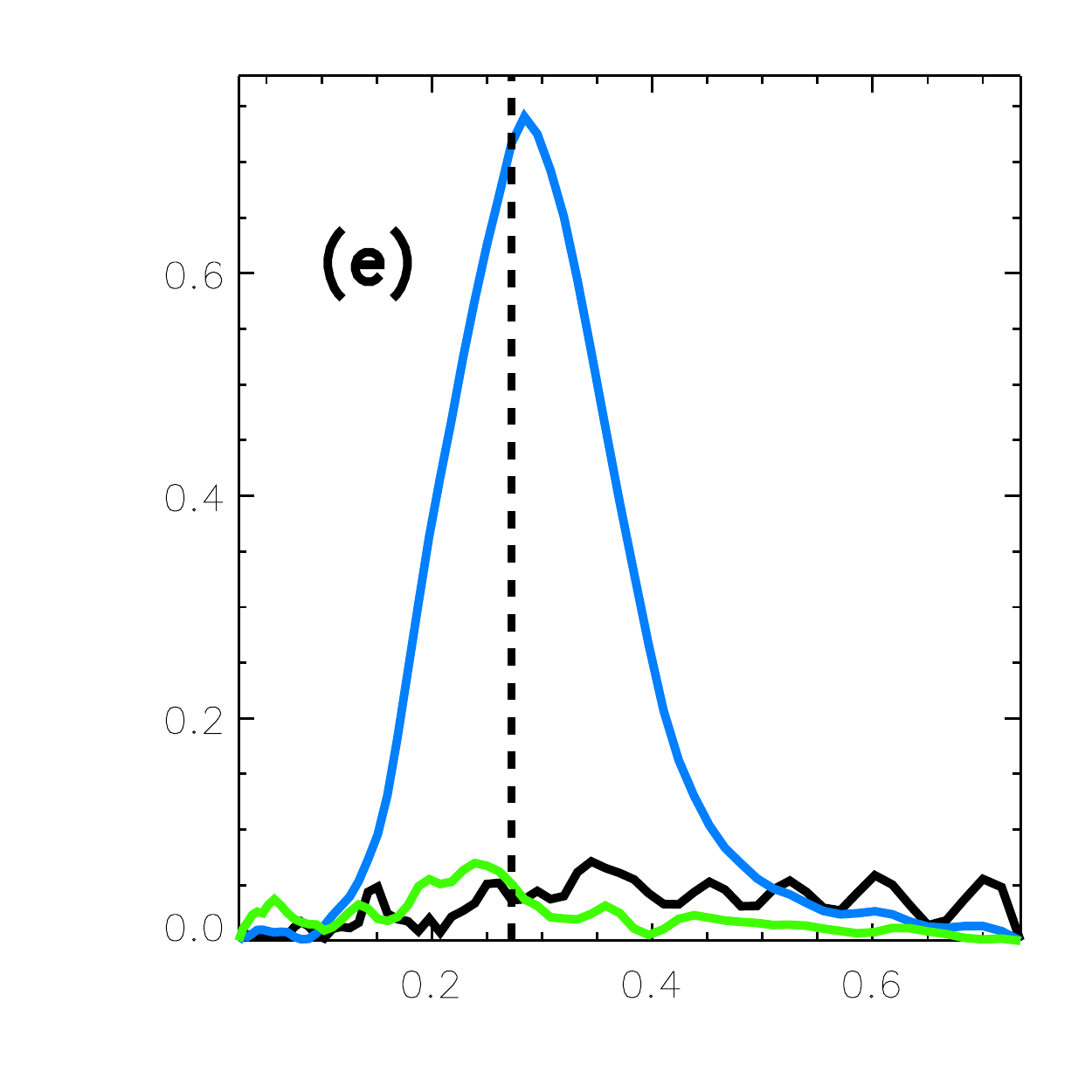}
  \includegraphics[width=0.15\textwidth]{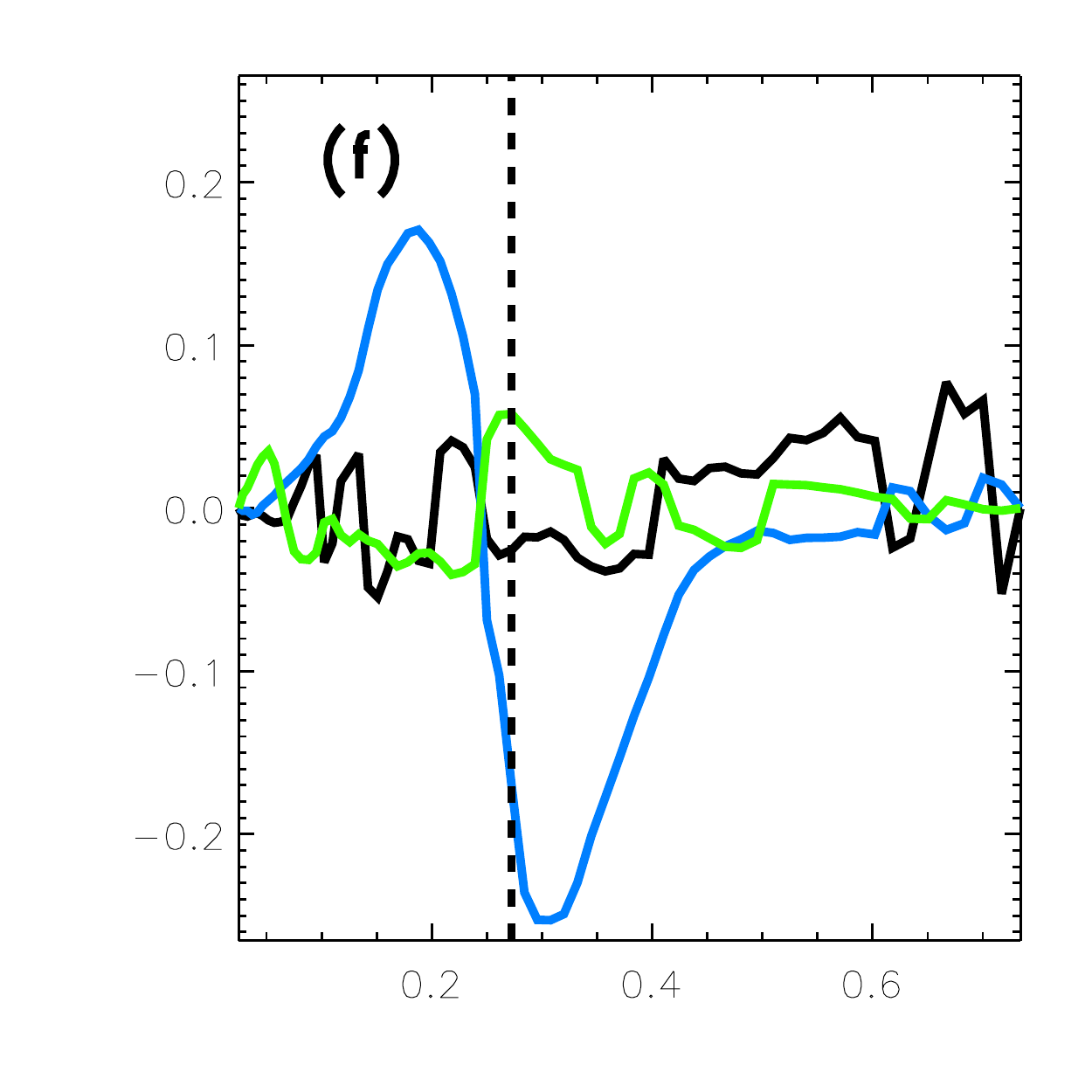}
  \includegraphics[width=0.15\textwidth]{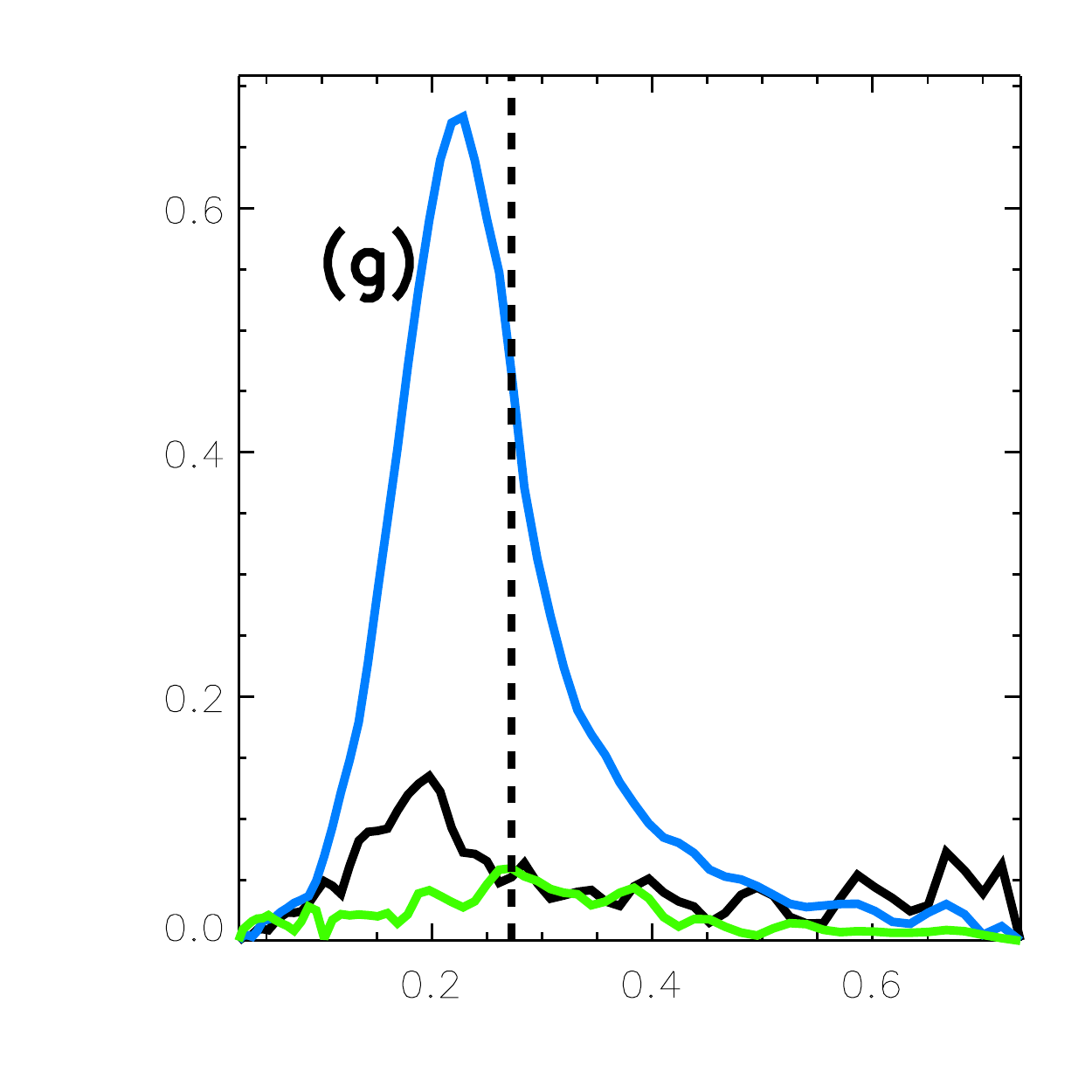}
  \includegraphics[width=0.15\textwidth]{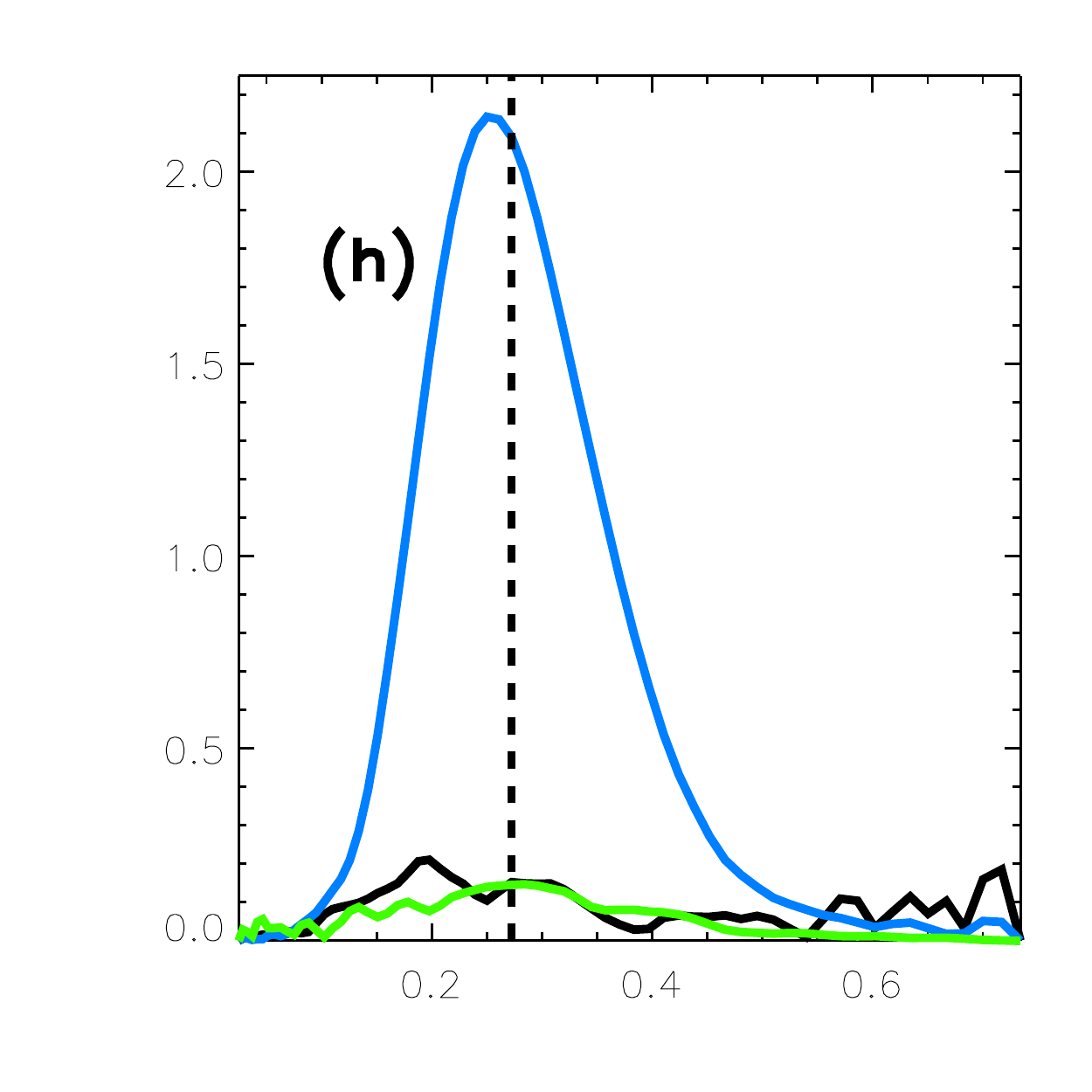}
\caption{Radial profiles of electrostatic potential $e\phi/T_e$.
The black, blue and green lines are corresponding to the poloidal mode number $m=5,6,7$, respectively.
The dotted line is the $\psi$ value at the $q=2$ rational surface.
The $x$ axis is $\psi/\psi_w$ and the $y$ axis is $e\phi/T_e$.
Time steps $(b)-(h)$ are corresponding to $(B)-(H)$ in \cref{fig:wavelet}. }\label{fig:phi-pol}
\end{figure}

\noindent
\\
\textbf{\large 2.} \textbf{\large Simulation} \textbf{\large parameters}
\\

This electron BAE is chosen to simplify our 2D nonlinear dynamical model,
since only the toroidal precessional resonance of deeply trapped particles is responsible\cite{jcheng2016} for the excitation.
In our simulations of a tokamak with a concentric circular cross-section,
the safety factor profile is
$q=1.797+0.8(\psi/\psi_w)-0.2(\psi/\psi_w)^2$
so that the $q=2$ rational surface is located at a minor radius $r_s=0.5a$,
where $\psi$ is the poloidal flux
with $\psi=0$ on the axis and $\psi=\psi_w$ at the plasma boundary,
and $a$ is the tokamak minor radius at the wall.
The inverse aspect ratio is $\epsilon\equiv a/R_0=0.333$
in terms of on-axis major radius $R_0$.
The thermal plasma temperature is uniform with $T_i=T_e$,
and the on-axis beta is $\beta=4\pi n_0(T_e+T_i)/B_0^2=0.00718$ with $B_0$ being the on-axis magnetic field.
The thermal electron density $n_{0}$ is uniform and
the energetic electron density profile is
$n_f=0.05n_0(1.0+0.20(\tanh((0.24-\psi/\psi_w)/0.06)-1.0))$,
so that the energetic electron density gradient peaks near the $q=2$ flux surface with $R_0/L_{n_f}=15$,
where $L_{n_f}\equiv -(d \ln n_f/dr)^{-1}$ is the density gradient scale length of energetic electrons.
The thermal ion density $n_{0i}$ is obtained through the neutrality condition $n_{0i}=n_{0}+n_f$.
The energetic electrons are loaded with a local Maxwellian distribution with the uniform temperature $T_f=25T_e$.
\\
\noindent
\\
\textbf{\large 3.} \textbf{\large Fast and repetitive frequency chirping of BAE.}
\\

In the linear simulations\cite{jcheng2016},
the most unstable toroidal mode number is $n=3$ with
$k_\theta\rho_i=0.115$ at the $q=2$ rational surface,
where the device size is $a=104\rho_i$,
$k_\theta=nq/r$ is the poloidal wave-vector,
$\rho_i=\sqrt{m_iT_i}/eB_0$ is the thermal ion gyro-radius,
and $m_i$ is the ion mass.
Given the thermal plasma profiles,
the frequency at the BAE accumulation point is $\omega_0=\sqrt{11T_i/2m_iR_0^2}\approx 2.34v_i/R_0$,
where $v_i=\sqrt{T_i/m_i}$ is the ion thermal velocity.
In our collisionless simulations, particle sources or sinks were turned off to allow free evolution of temperature and density profiles.

The nonlinear time evolution of the electrostatic potential of the $(n, m)=(3, 6)$ mode is shown in \cref{fig:wavelet},
where the amplitude (red line in panel a) and spectrum (panel b) of the electrostatic potential are measured at the mode rational surface $r_s$. 
The time evolution of the power spectrum in the lower panel is obtained through the wavelet analysis of the real part of the electrostatic potential $\phi$.
\cref{fig:wavelet}b shows a fast oscillation of the mode intensity.
For each time step in \cref{fig:wavelet}b,
the frequency of the maximum power intensity is plotted in \cref{fig:wavelet}a with a black curve,
which shows
that the frequency starts with a linear frequency of $\omega_{BAE}$ and chirps in the nonlinear stage roughly in phase with the intensity of the electrostatic potential $\phi$ (red curve).
In the linear stage, the measured real frequency is
$\omega_{BAE}=1.215\omega_0$ (the straight black line before point C) and the linear growth rate $\gamma=0.0341\omega_0$,
The mode saturates at point D with the amplitude of the perturbed electrostatic potential being $|e\phi/T_e|=6.84\times 10^{-3}$.
After saturation (point D), the mode evolves to a nonlinear chirping state.
The first chirping period (points F-I) is about seven wave periods, which is in a non-adiabatic regime\cite{zonca2015nonlinear}.
There is a small phase difference between the minima of amplitude and frequency.
During such chirping, the range of the frequency change is $0.157\omega_0$,
which is $12.9 \%$ of the linear eigenfrequency $\omega_{BAE}$.

\cref{fig:phi-snap} and \cref{fig:phi-pol} shows the time evolution of the BAE mode structure of the electrostatic potential on a poloidal plane and in the radial direction.
The mode structure is coherent and nearly symmetric with respect to the $q=2$ rational surface.
It changes little from the linear stage (panel b) to the nonlinear saturation with a maximal mode amplitude (panel d).
It then moves slightly outward as the mode amplitude decreases and the frequency starts chirping down (panel e).
At the lowest amplitude and frequency (panel f), the mode structure is destroyed across the $q=2$ rational surface.
Later, a coherent mode structure forms but moves slightly inward as the mode amplitude increases and the frequency chirps up (panel g).
Finally, at the highest mode amplitude and frequency (panel h), the mode structure becomes nearly symmetric with respect to the $q=2$ rational surface,
similar to that in the initial nonlinear saturation (panel d). This time oscillation (from panel d to panel h) for the mode amplitude, frequency, and mode structure repeats itself in the next cycle (from panel h to panel j) but with a slightly smaller time period.

In controlled simulations, we find no frequency chirping when the nonlinearity of the energetic electrons is removed.
This indicates that energetic electrons are exclusively responsible for such a frequency chirping process.
The saturation amplitude significantly increases when the nonlinearity of thermal ions is removed, which suggests that thermal ions contribute to the nonlinear saturation.

\begin{figure}[!hpt]
   \includegraphics[width=0.23\textwidth,height=0.15\textwidth]{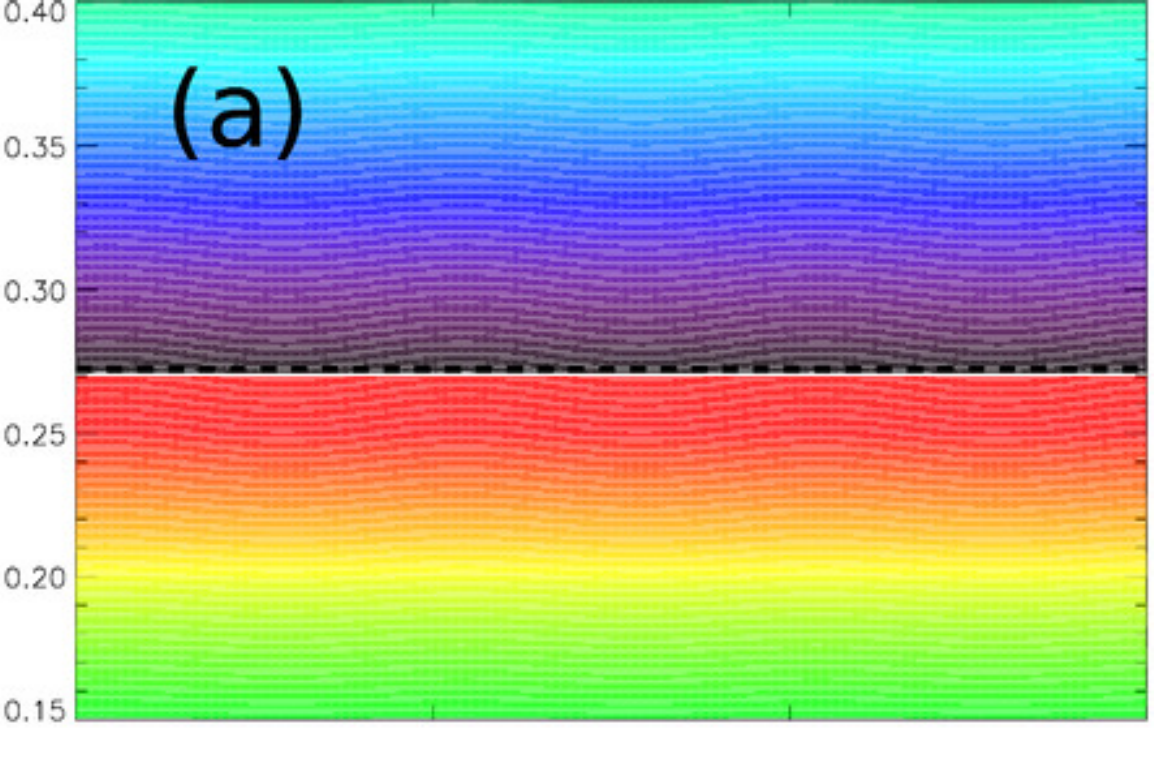}
   \includegraphics[width=0.23\textwidth,height=0.15\textwidth]{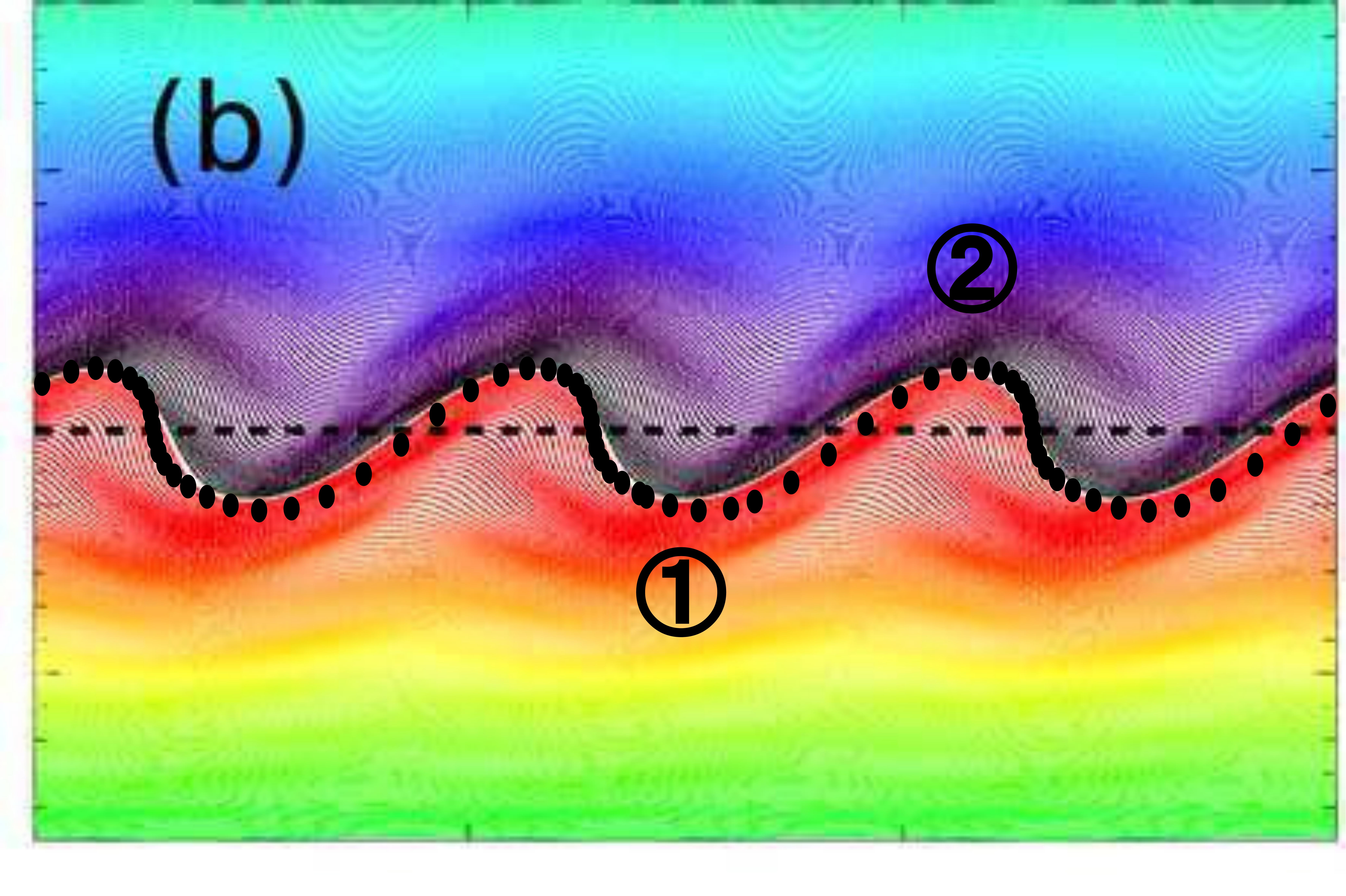}
   \includegraphics[width=0.23\textwidth,height=0.15\textwidth]{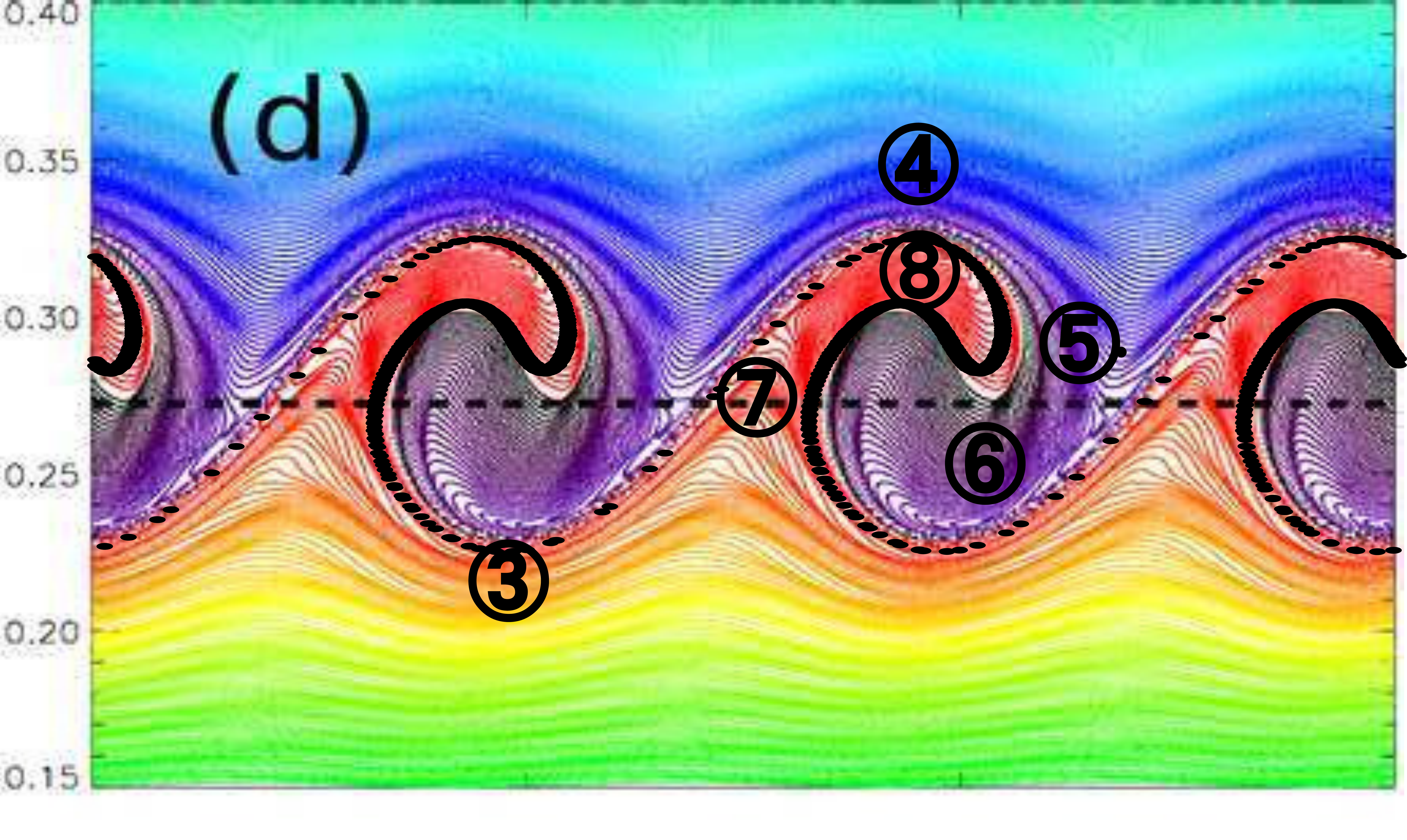}
   \includegraphics[width=0.23\textwidth,height=0.15\textwidth]{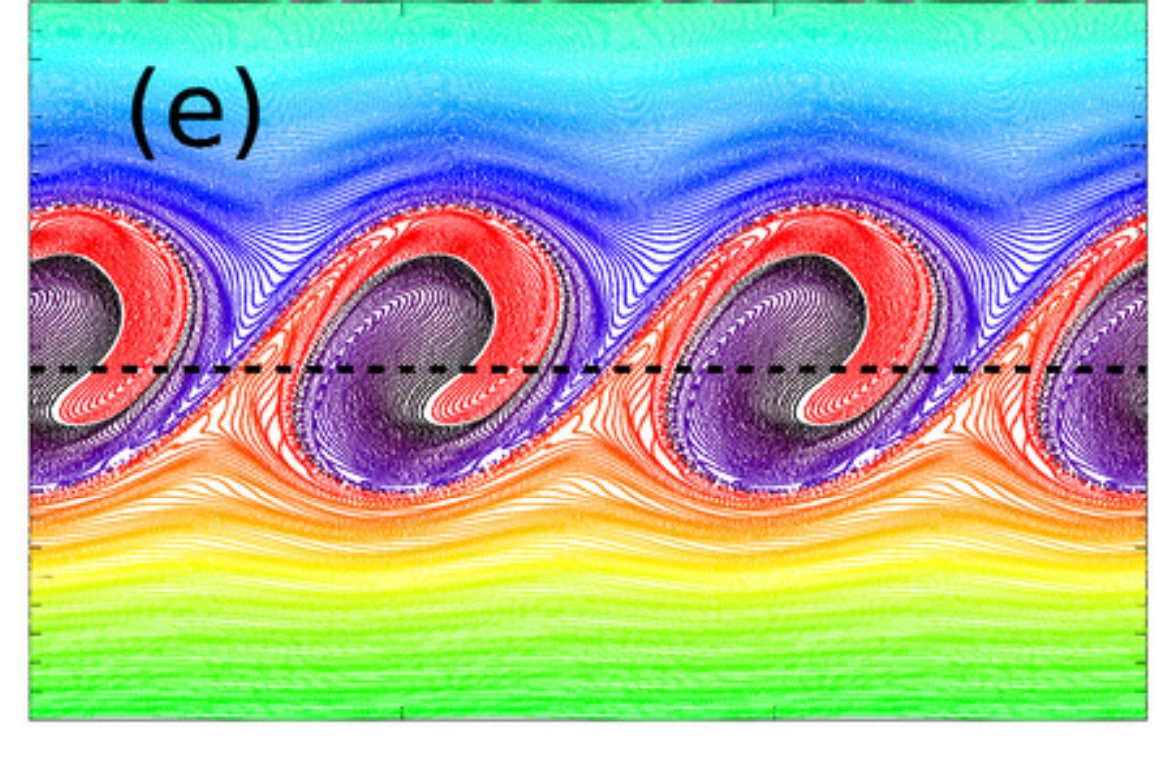}
   \includegraphics[width=0.23\textwidth,height=0.15\textwidth]{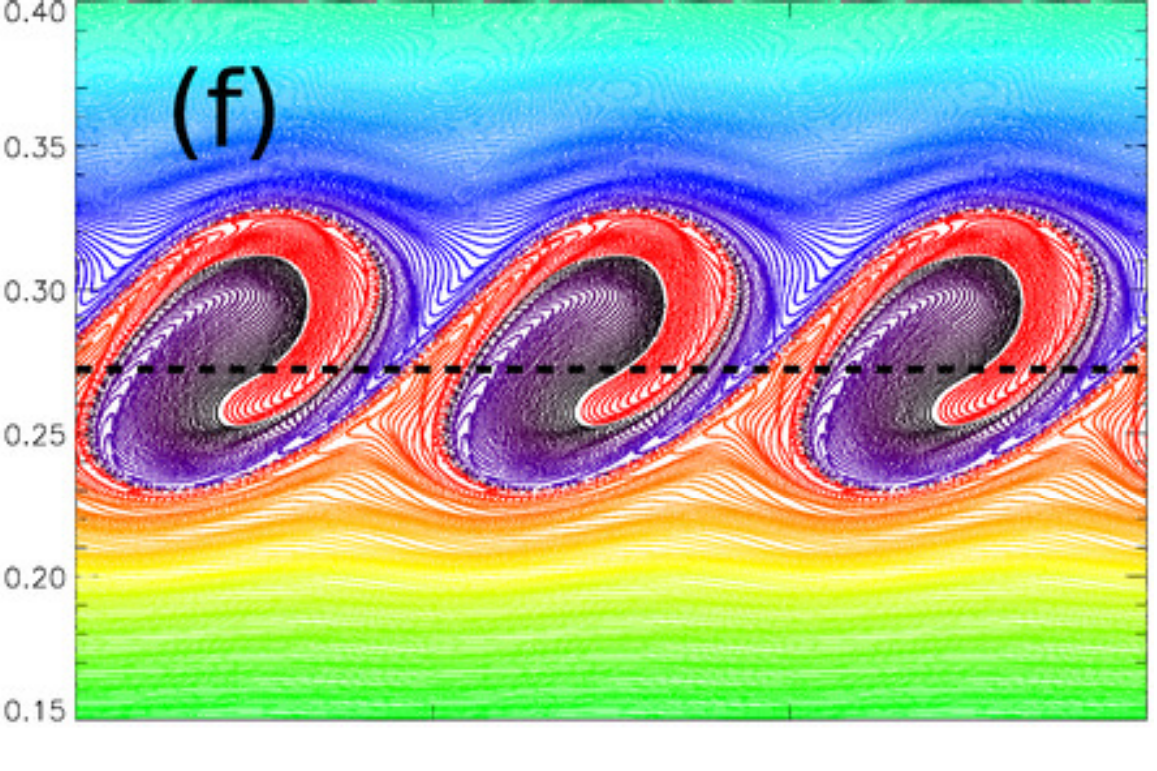}
   \includegraphics[width=0.23\textwidth,height=0.15\textwidth]{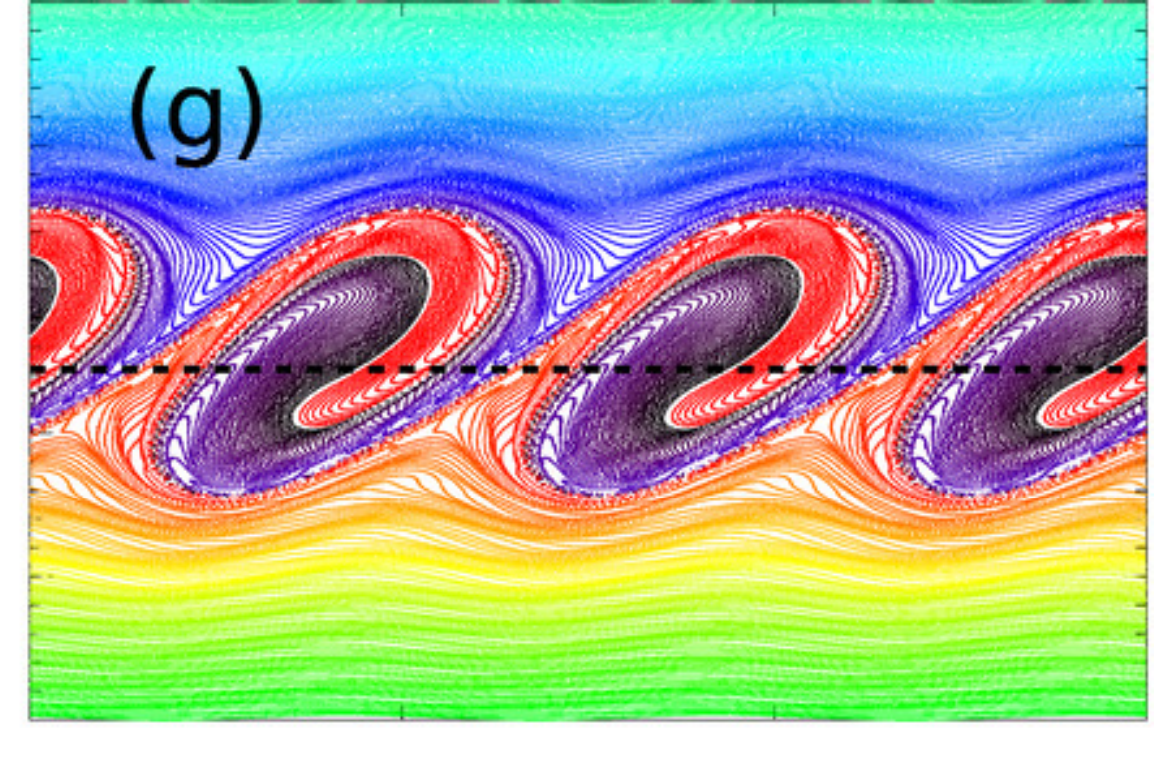}
   \includegraphics[width=0.23\textwidth,height=0.15\textwidth]{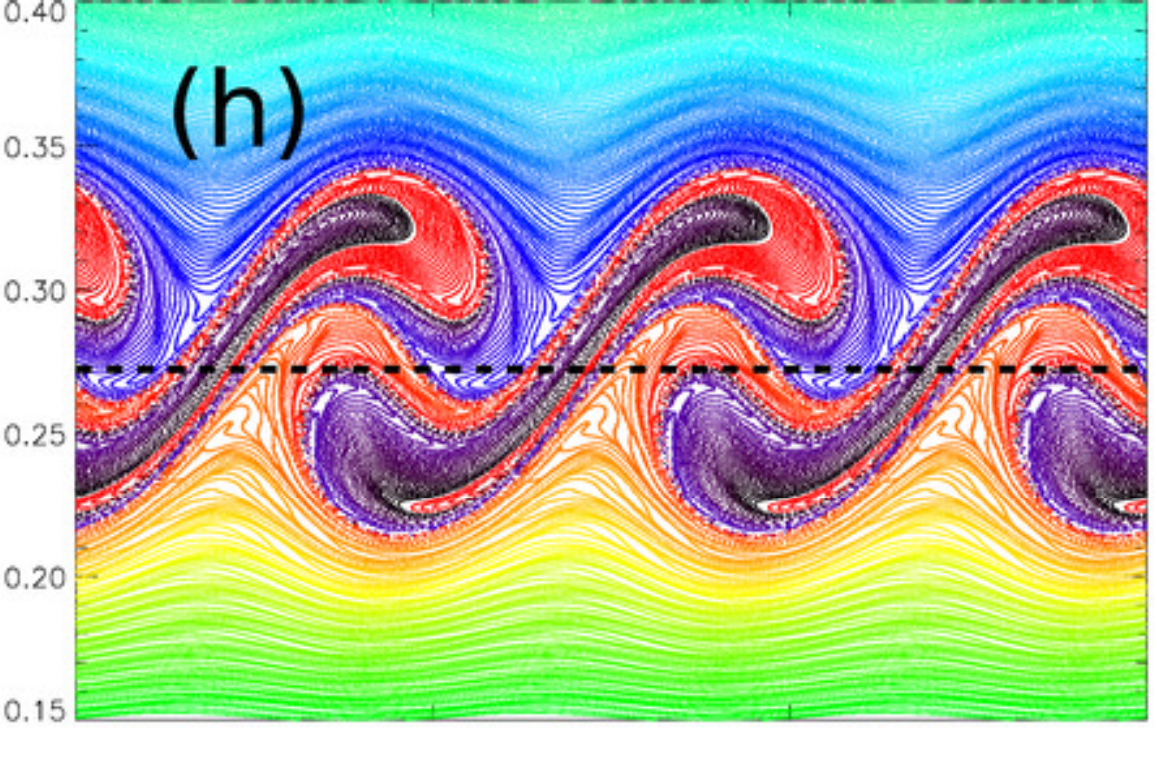}
   \includegraphics[width=0.23\textwidth,height=0.15\textwidth]{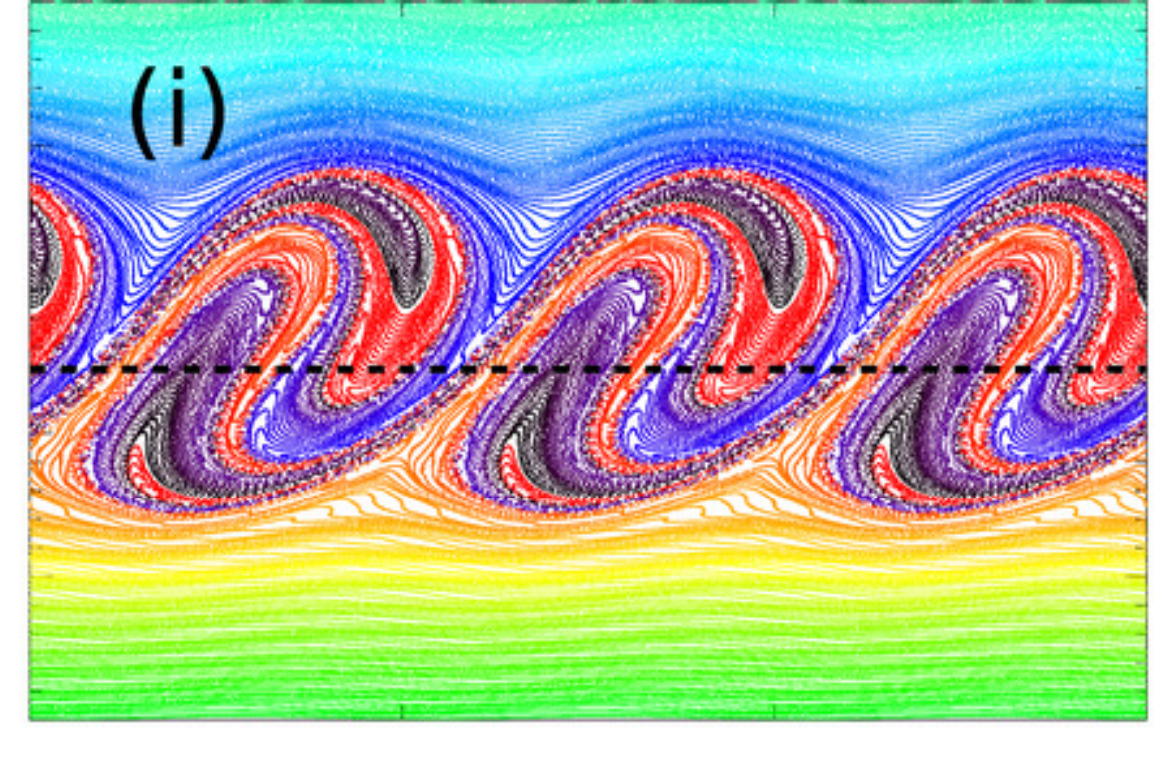}
   \includegraphics[width=0.23\textwidth,height=0.15\textwidth]{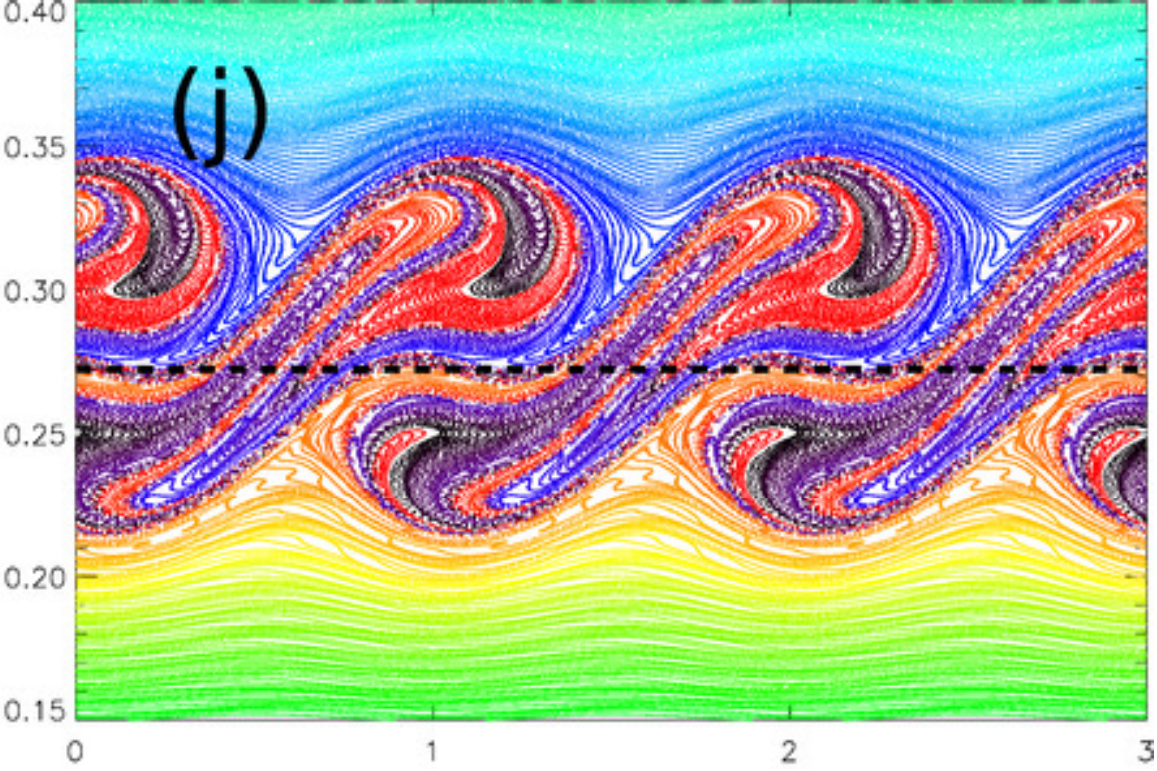}
   \includegraphics[width=0.23\textwidth,height=0.15\textwidth]{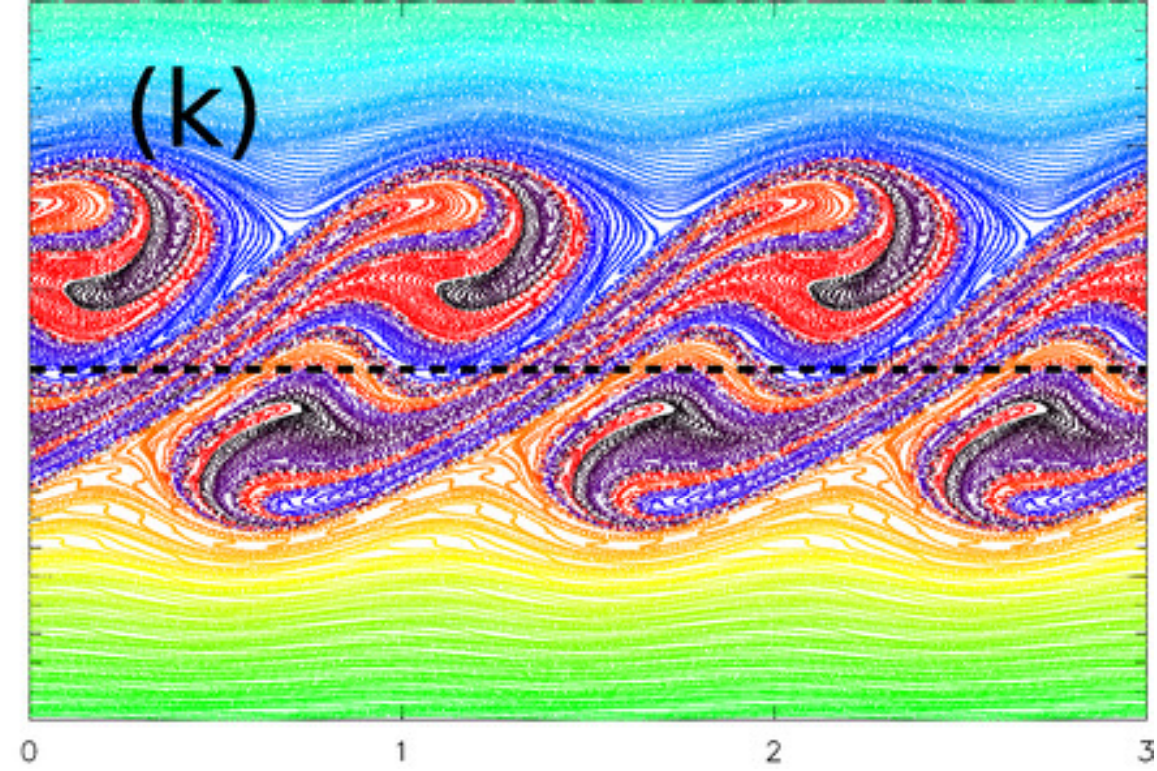}
   \caption{Time evolution of passive energetic-particle trajectories in $(\delta \zeta, P_\zeta)$ Lagrangian space.
Different color represents initial canonical angular momentum $P_\zeta$ as shown in panel (a).
Dashed line is the $q=2$ rational surface.
Horizontal axis is the toroidal angle $\delta\zeta/2\pi$ in the wave frame, and vertical axis is $-P_\zeta/\psi_w$.
The diagram of predicted particle trajectories according to \cref{pzdot-2} are labeled in panel (b) with black dot line, respectively.
Labels $a-k$ correspond to $A-K$ in \cref{fig:wavelet}.}\label{fig:tracking-pzeta}
\end{figure}

\begin{figure}[!ht]
   \includegraphics[width=0.23\textwidth,height=0.15\textwidth,trim=0 20 0 0, clip]{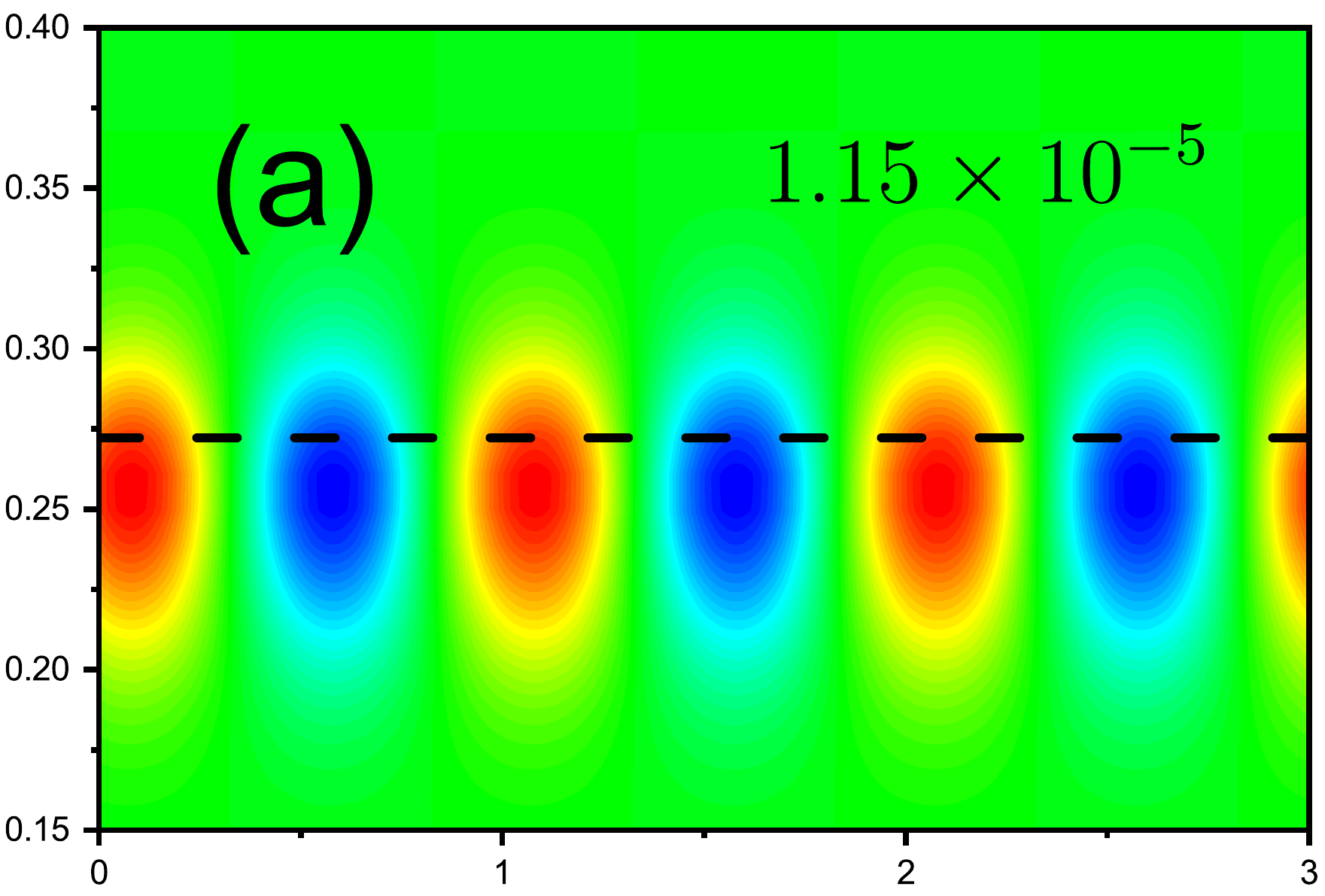}
   \includegraphics[width=0.23\textwidth,height=0.15\textwidth,trim=30 20 0 0, clip]{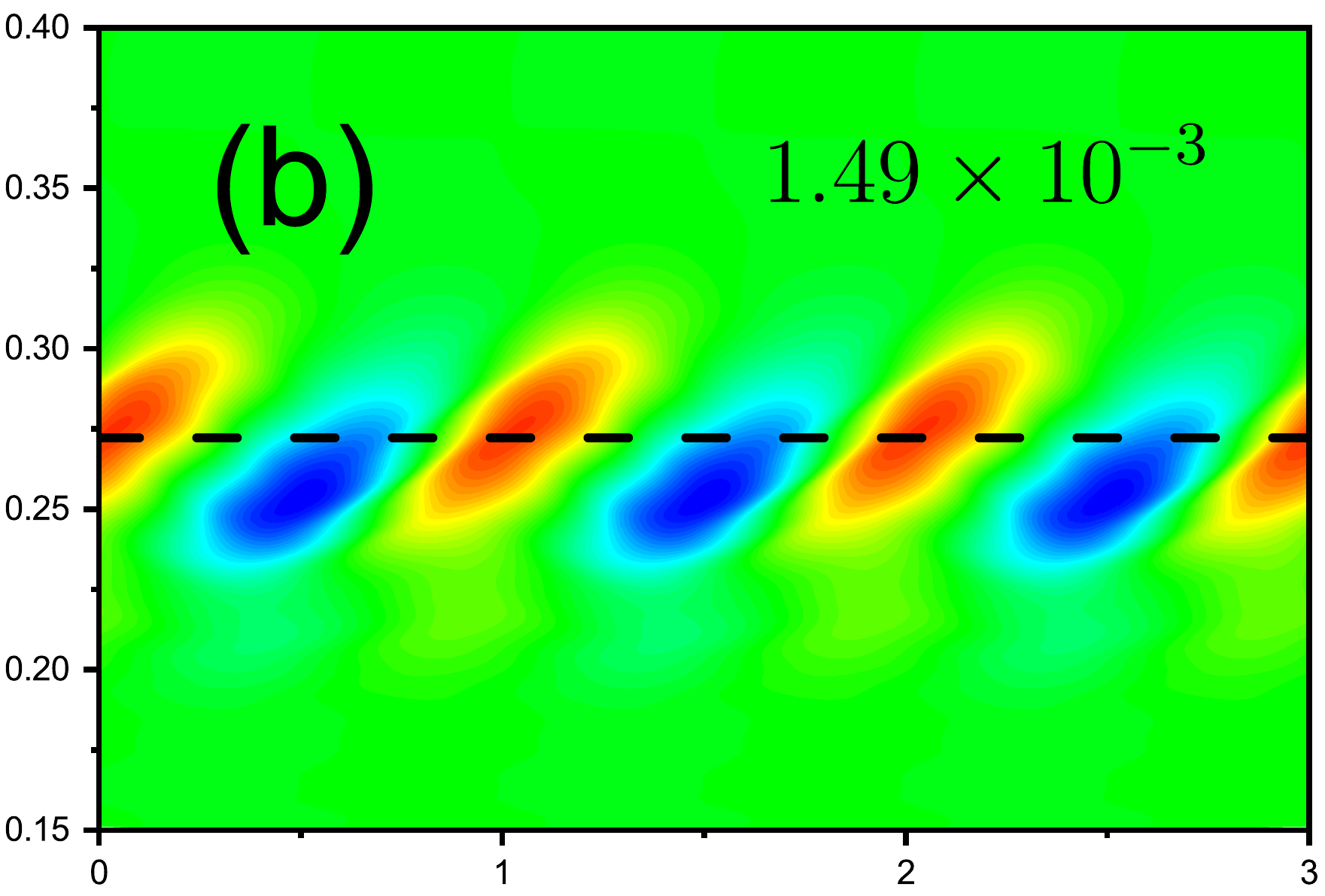}
   \includegraphics[width=0.23\textwidth,height=0.15\textwidth,,trim=0 20 0 0, clip]{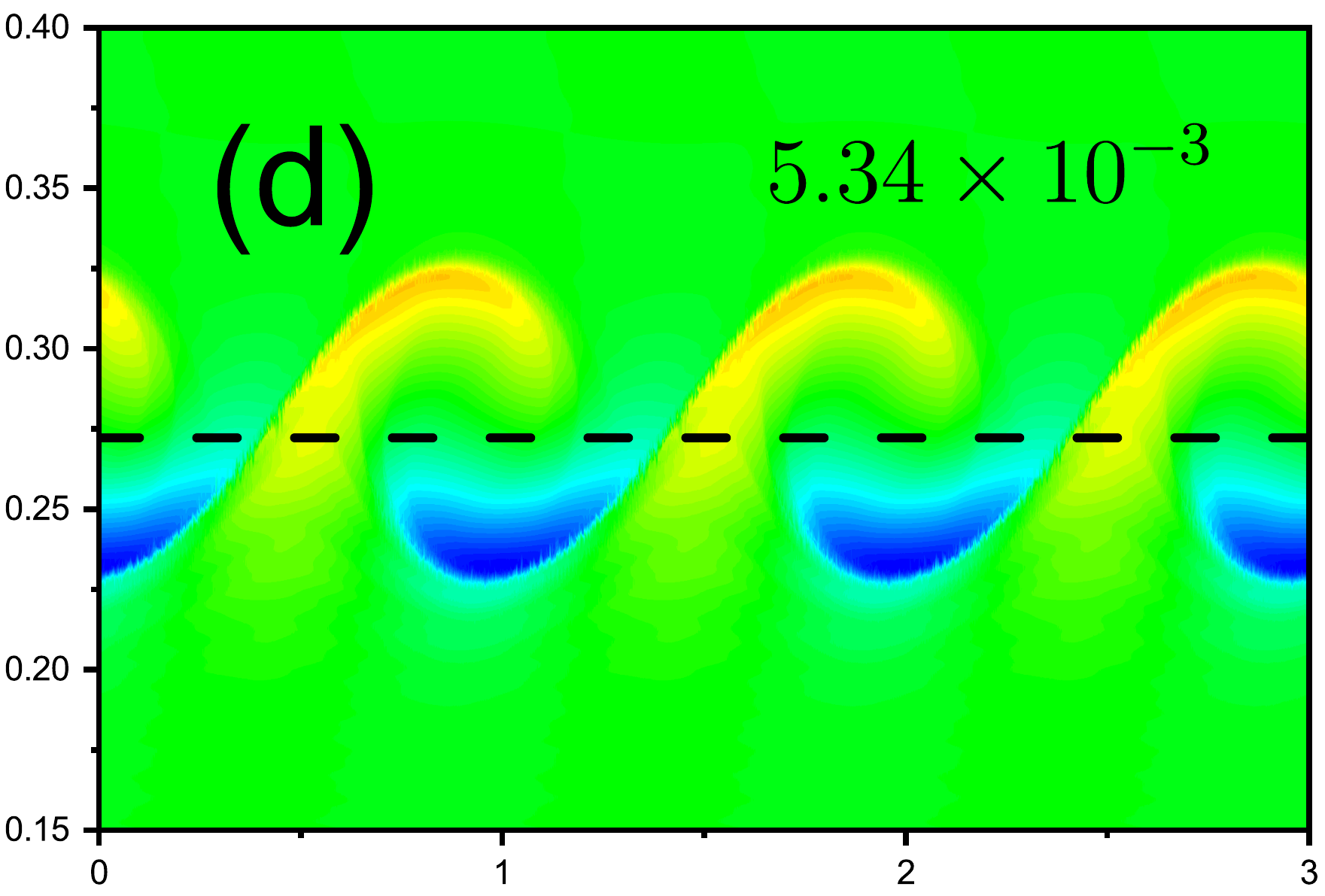}
   \includegraphics[width=0.23\textwidth,height=0.15\textwidth,trim=30 20 0 0, clip]{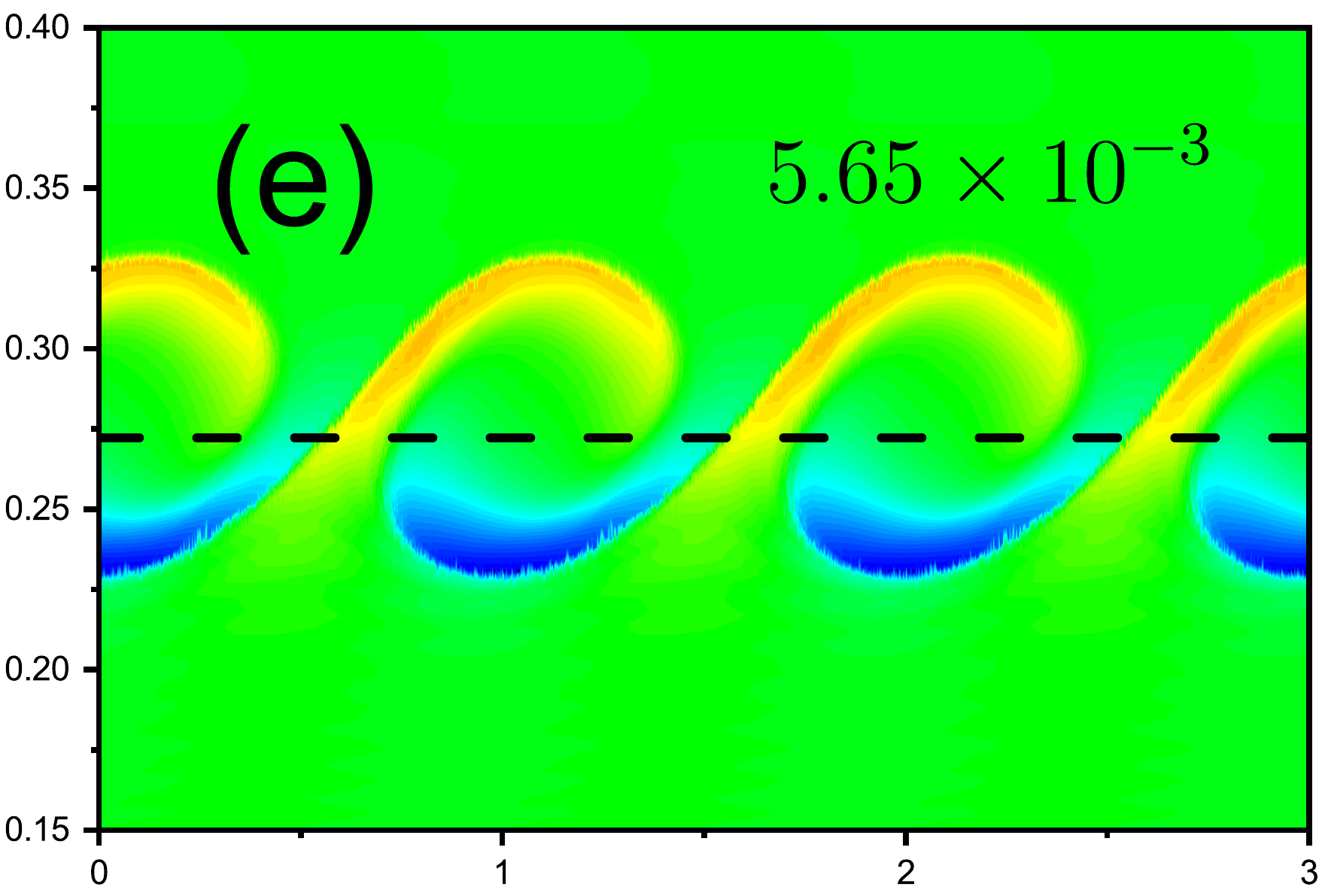}
   \includegraphics[width=0.23\textwidth,height=0.15\textwidth,trim=0 20 0 0, clip]{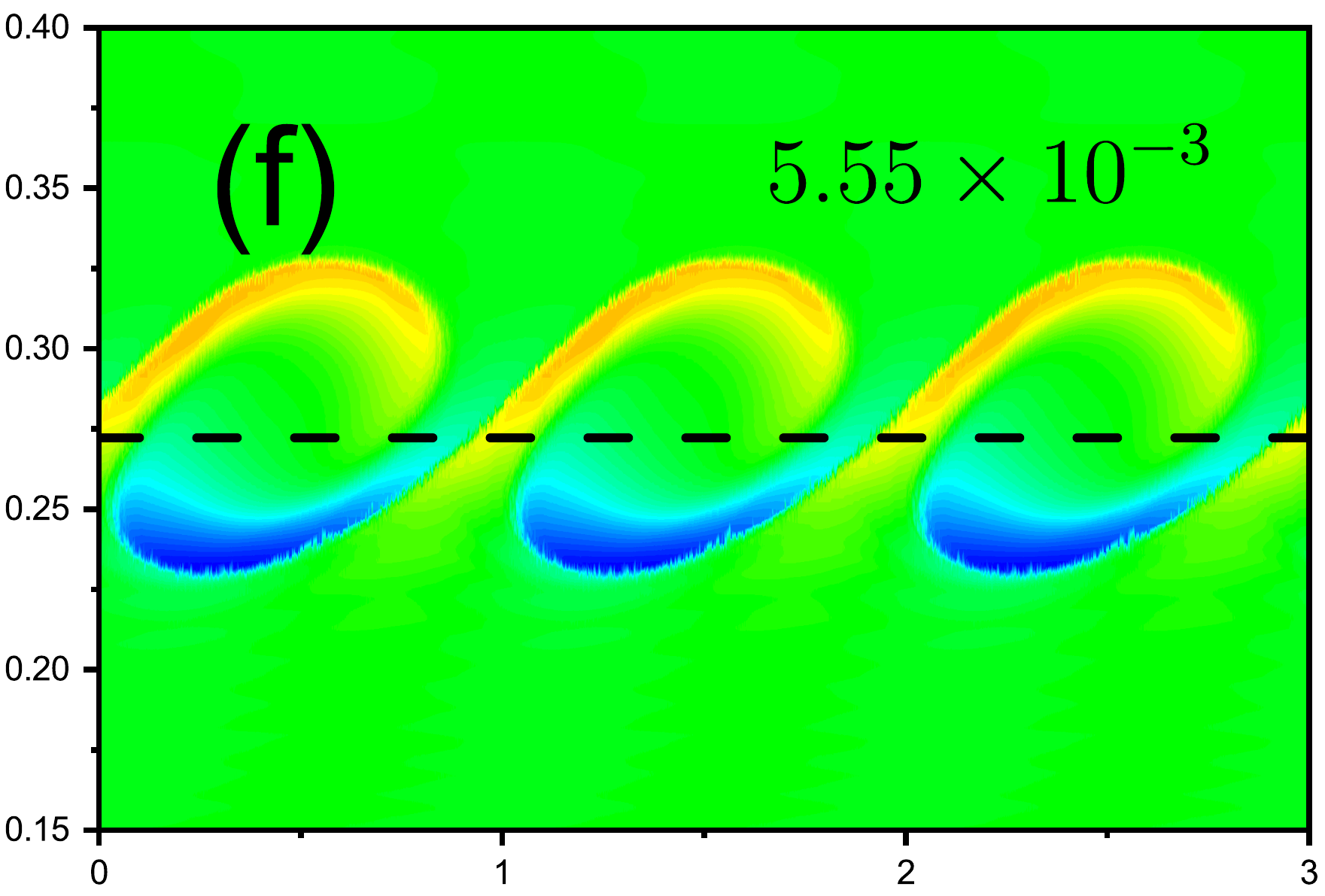}
   \includegraphics[width=0.23\textwidth,height=0.15\textwidth,trim=30 20 0 0, clip]{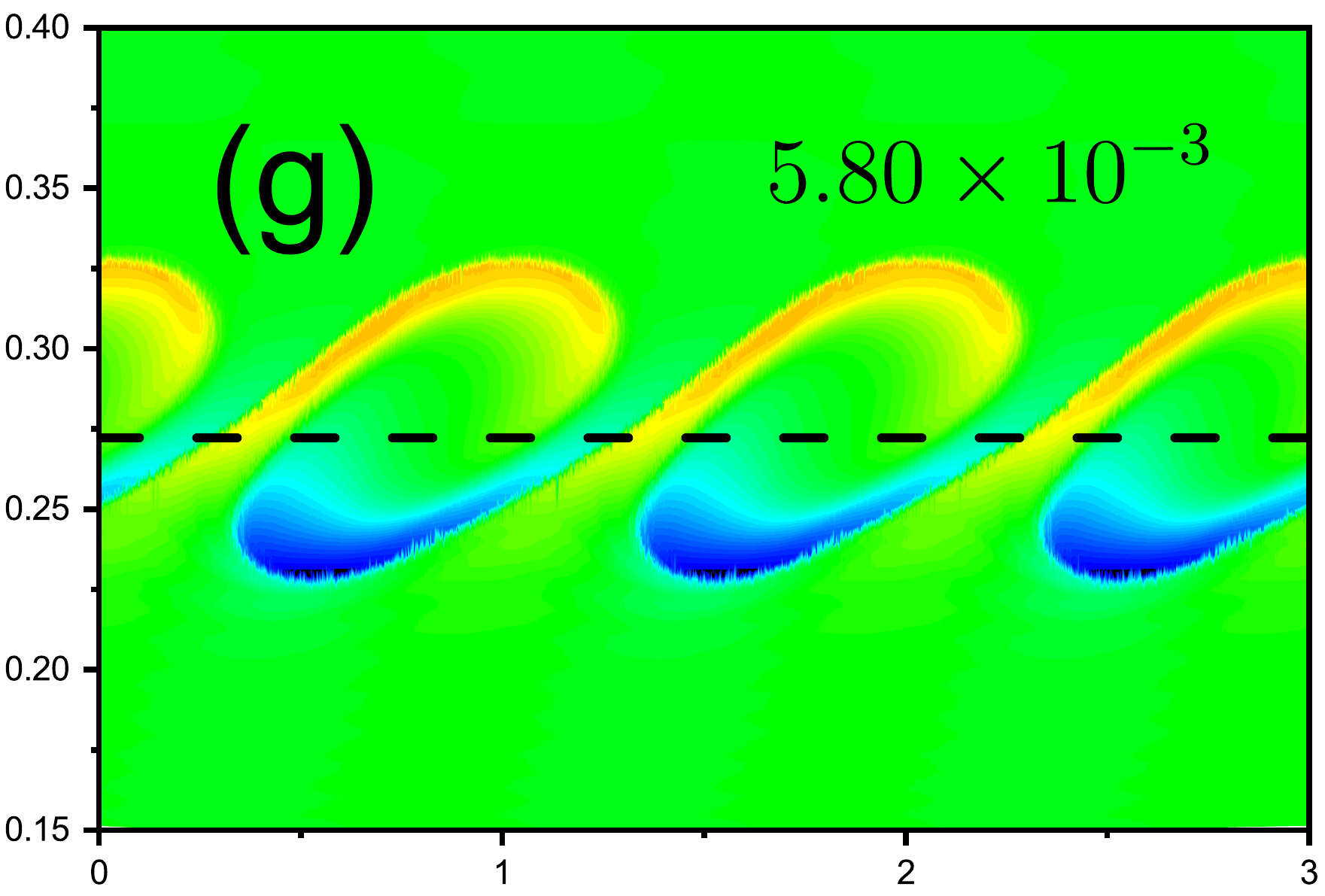}
   \includegraphics[width=0.23\textwidth,height=0.15\textwidth,trim=0 20 0 0, clip]{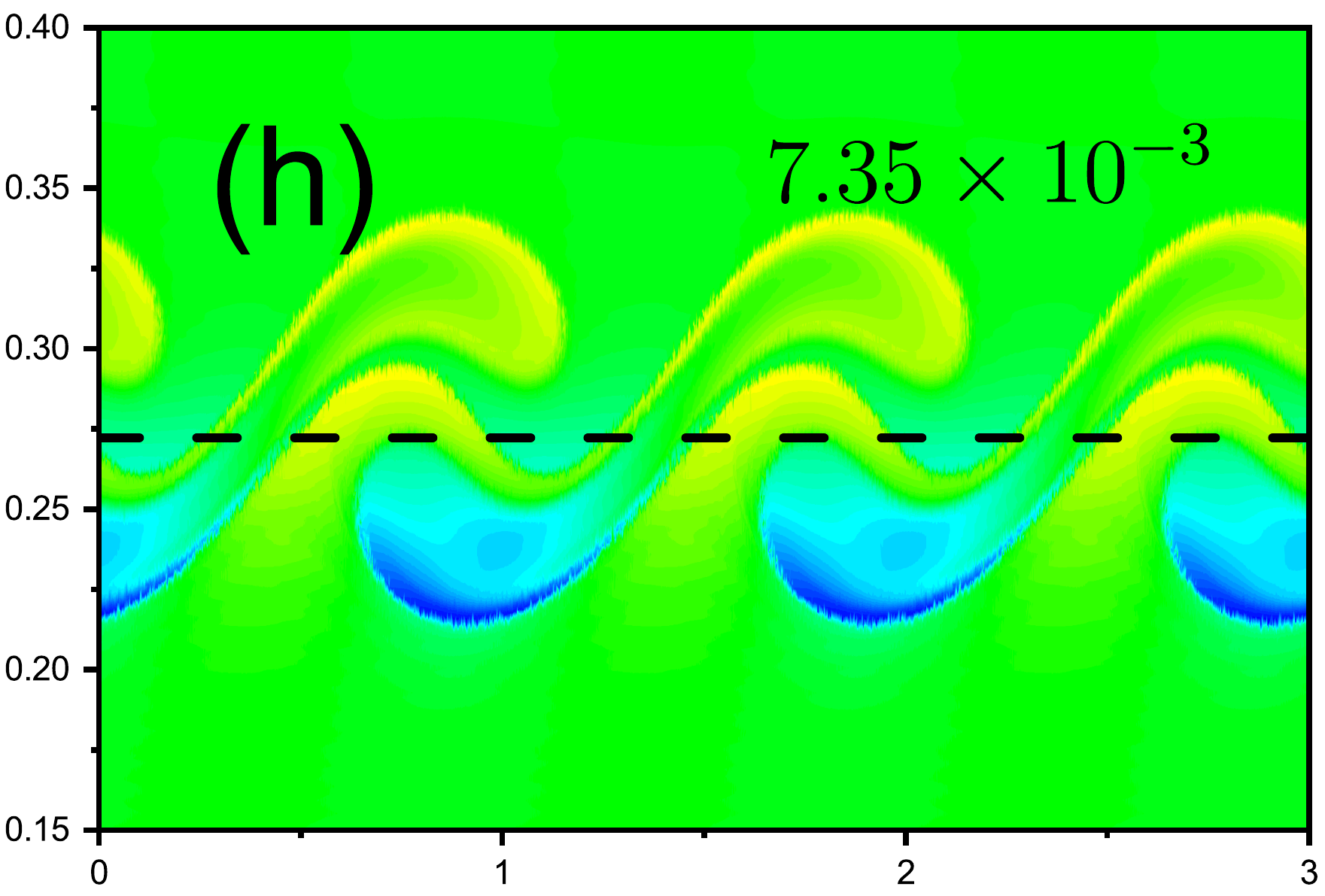}
   \includegraphics[width=0.23\textwidth,height=0.15\textwidth,trim=30 20 0 0, clip]{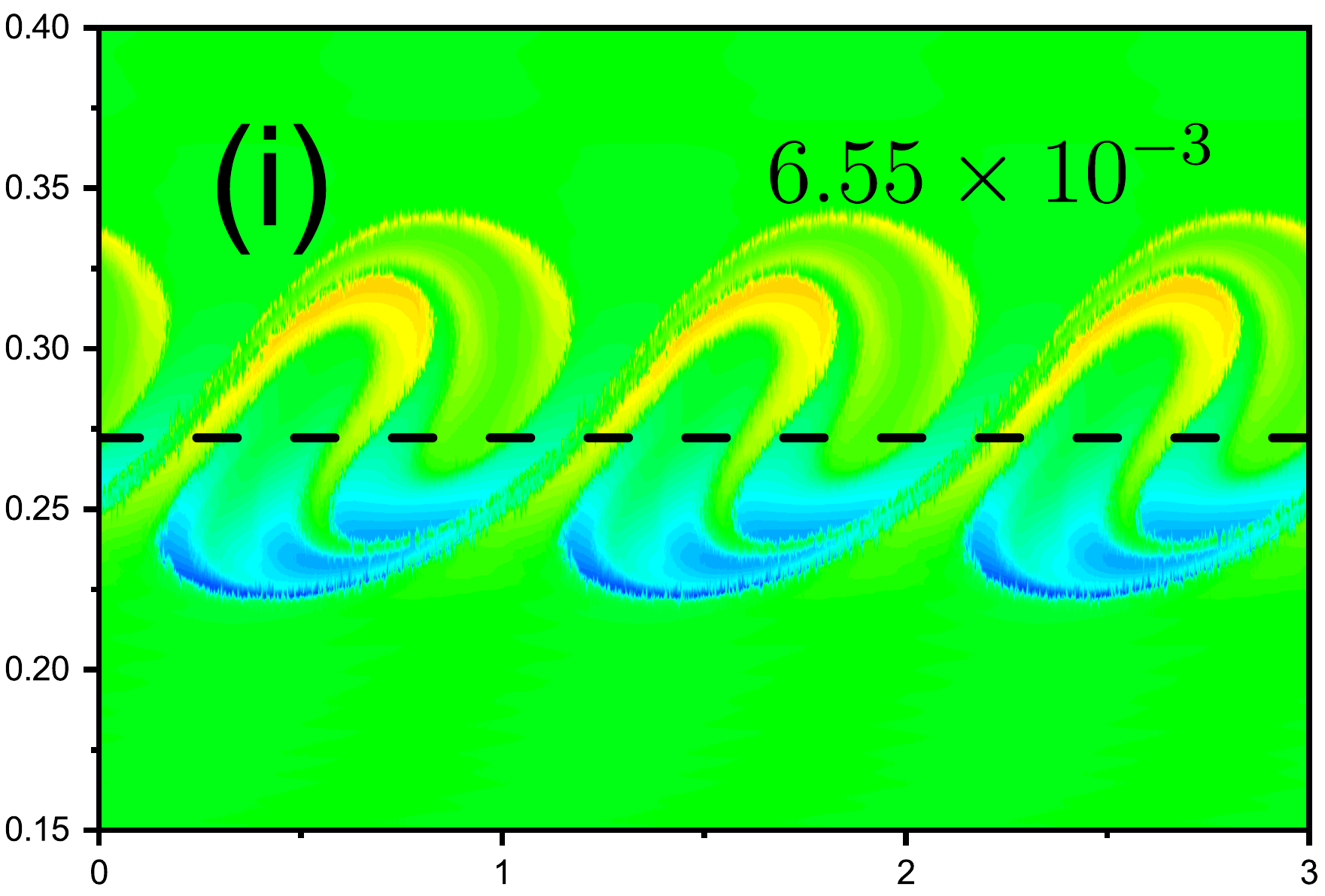}
   \includegraphics[width=0.23\textwidth,height=0.15\textwidth,trim=0 0 0 0, clip]{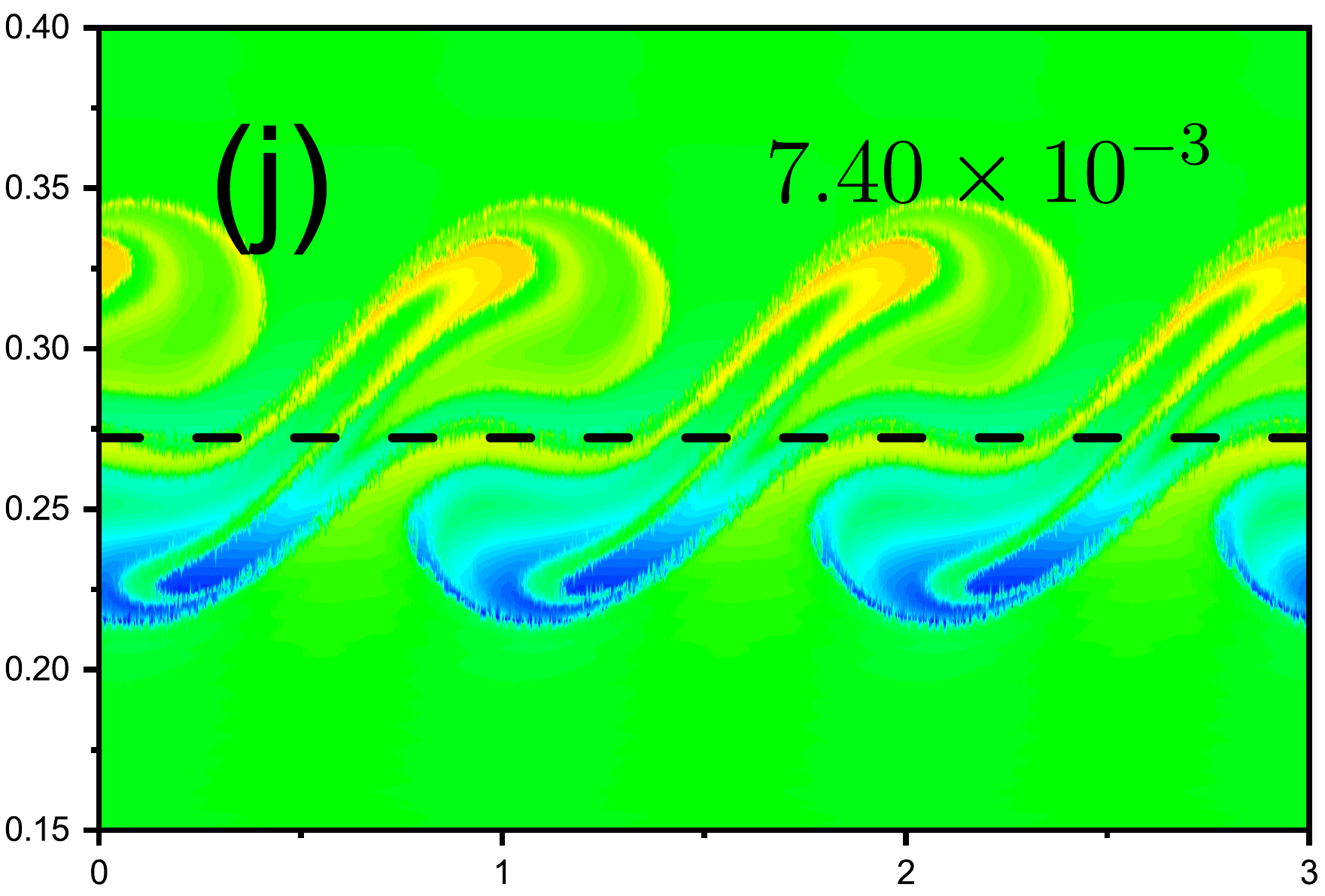}
   \includegraphics[width=0.23\textwidth,height=0.15\textwidth,trim=30 0 0 0, clip]{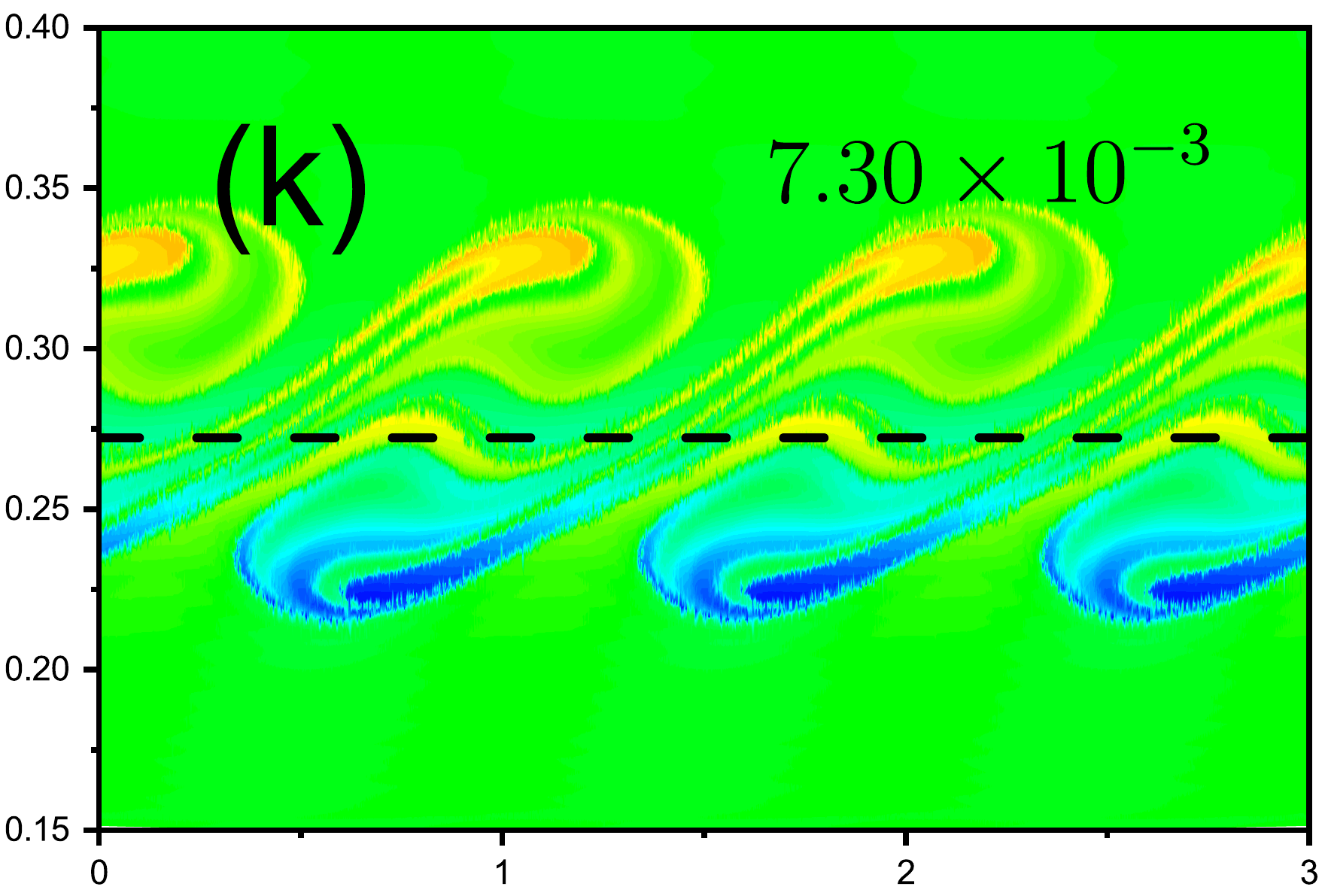}
   \caption{Evolution of passive energetic-particle perturbed distribution function $\delta f(\delta\zeta, P_\zeta)$ in Eulerian phase space.
Different color represents amplitude of perturbed distribution function $\delta f/f$.
Dashed line is the $q=2$ rational surface.
Horizontal axis is the toroidal angle $\delta\zeta/2\pi$ in the wave frame, and vertical axis is $-P_\zeta/\psi_w$.
The maximum of the absolute perturbed distribution $|\delta f/f|$ for each time point is indicated.
Labels $a-k$ correspond to $A-K$ in \cref{fig:wavelet}.}\label{fig:deltf-pzeta}
\end{figure}

\noindent
\\
\textbf{\large 4.} \textbf{\large Intrinsically 2D particle dynamics.}
\\

Given an energy $E$ and a magnetic moment $\mu$, the guiding-center motions of the deeply trapped particles can be described by a pair of action-angle variables $(\zeta,
P_\zeta)$ with $\zeta$ being the toroidal angle, $P_\zeta=g\rho_\|-\psi$ being the normalized canonical angular momentum \cite{white2013theory}, $2\pi g$ being the poloidal current and
$\rho_\|=v_\|/B$ being the normalized parallel velocity.
Here, we examine the structure of the energetic particle distribution function in the $(\zeta, P_\zeta)$ phase space.
Instead of directly analyzing the data of energetic particles in the self-consistent simulations, 
tracking orbits of passive particles will give a better resolution and a lower particle noise for diagnosis by loading a large number of the passive deeply-trapped particles in the resonant region of the phase space.

These passive particles are uniformly initialized in the $(\psi,\zeta)$ plane
with $\theta=0$, $v_\|=-0.01\sqrt{T_e/m_e}$ and $\mu=32T_e/B_0$ within the resonant region.
$P_\zeta / \psi_w$ are initialized in the range of $(0.15, 0.42)$.
The particle positions in \cref{fig:tracking-pzeta} are plotted in the $(\zeta, P_\zeta)$ space, where $P_\zeta$ values are approximately the radial position since $P_\zeta\approx-\psi$ for the energetic electrons. 

Using the nonlinear gyrokinetic equations \cite{brizard2007foundations, pliu2017a, pliu2017b, holod2009electromagnetic}, we can calculate the trajectory of the deeply trapped particles using the perturbed electrostatic field $\phi=\hat{\phi}(\psi,\epsilon t)e^{i \alpha}$, where $\epsilon$ indicates the slow evolution of the mode amplitude and $\alpha = n\zeta-m\theta-\omega t$, $\alpha\in [0,2\pi)$, with a single toroidal mode number $n$. Since the parallel wave wavelength is much longer than the distance that a deeply trapped particle travels along the magnetic field line, i.e., $k_\|\approx 0$, the guiding center equation of motion is
\begin{small}
\begin{equation}\label{pzdot-2}
\dot{\zeta} \propto \frac{\mu}{e}\frac{\partial B_0}{\partial\psi} - \frac{\partial\hat{\phi}}{\partial\psi} e^{i\alpha}, \qquad 
\dot{\psi} \propto n\hat{\phi}e^{i(\alpha +\pi/2)},
\end{equation}
\end{small}%
where $e$ is the elementary charge. Since in \cref{pzdot-2}, the second term of $\dot{\zeta}$ and $\dot{\psi}$ have a $\pi/2$ phase difference, 
these two equations describe a Lissajous curve \cite{greenslade1993all} rotating clockwise with a distortion described by the first term of $\dot{\zeta}$. A 2D dynamical model can be constructed using the reduced evolution equation for the perturbed distribution function,
\begin{small} 
\begin{equation}\label{dfdt}
\dot{\delta f} \propto nf_0\left(\frac{\mu}{e}\frac{\partial B_0}{\partial\psi}+\frac{1}{f_0}\left.\frac{\partial f_0}{\partial\psi}\right|_{v_\perp}\right)\hat{\phi}e^{i(\alpha+\pi/2)},
\end{equation}
\end{small}%
which is closed with the field equations.

By using Taylor expansion around the $q=2$ rational surface and keeping the first and second order terms of $B_0$ and eliminating the phase dependence for $\dot{\zeta}$, and keeping the zero order terms of $\hat{\phi}$ for $\dot{\psi}$, the equations are reduced to
\begin{small}
\begin{equation}\label{pzdot-4}
\dot{\delta\zeta} \propto \frac{\mu}{e} \frac{\partial^2 B_0}{\partial\psi^2}\bigg|_{\psi_0}\delta\psi, \qquad
\dot{\delta\psi} \propto n\hat{\phi}(\psi_0)e^{i(\alpha+\pi/2)},
\end{equation}
\end{small}%
where $\delta \zeta = \zeta - \omega_{d0}t$, $\delta\psi=\psi-\psi_0$, $\alpha = n\delta\zeta-m\theta-(\omega-n\omega_{d0})t$,
the toroidal precessional drift $\omega_d \propto (\mu/e) \partial B_0/\partial \psi\propto \mu B_0/r$,
$\omega_{d0}=\omega_d(\psi_0)$ and
$\psi_0$ is the poloidal flux at the $q=2$ rational surface.
Regarding $(\delta\zeta,\ \delta \psi)$ as the generalized canonical coordinates $(x,\ v)$, \cref{pzdot-4} recovers the conventional 1D nonlinear dynamical model\cite{o1971nonlinear}.

Compared with \cref{pzdot-2}, \cref{pzdot-4} neglects the asymmetric radial dependence of $B_0$ and $\hat{\phi}$ and the phase dependence for $\dot{\zeta}$. Thus, \cref{pzdot-2} generalizes the wave-particle resonant interactions to a non-perturbative and 2D dynamical process, which incorporates the radial
(perpendicular to the dominant wave-particle interaction)
dependence of the toroidal precessional drifts as well as the mode structure and the phase dependence (the second term of $\dot{\zeta}$ in \cref{pzdot-2}). In particular, the asymmetric radial dependence is essential to generate the nonlinear chirping phenomenon, which will be discussed later.

Using \cref{pzdot-2}, the energetic particle trajectories (\cref{fig:tracking-pzeta}) can be easily understood.
At time B, the trajectory starts twisting. The peak points \textcircled{1} and \textcircled{2} in the panel (b) are the $\dot{\delta\psi}=0$ turning points, corresponding to the $\alpha=0$ region.
At time D, the symmetric phase space island is formed.
The points \{\textcircled{3},\textcircled{4},\textcircled{6},\textcircled{8}\} in the panel (d) are the $\dot{\delta\psi}=0$ turning points with the particle velocity only along the negative and positive $\pmb {\delta\zeta}$ directions, corresponding to the $\alpha=0$ region.
The points of \textcircled{5} and \textcircled{7} are $\dot{\delta\zeta}=0$ turning points with the particle velocity only along the negative and positive $\pmb {\delta\psi}$ directions, 
corresponding to the $\alpha=3\pi/2+O$ and $\alpha=\pi/2-O$, where the corrections $O$ comes from the radial dependence of the toroidal precessional drift.

This paradigm is first verified with a simple numerical test, where the particle trajectories are modeled by \cref{pzdot-2} with all the parameters set to the actual data in the simulations. 
The curves were shown in \cref{fig:tracking-pzeta} (panel b, d, black dot line), which are the instantaneous interface of particles initially residing at the two sides of rational surface. Such a good agreement further supports the validity of our paradigm.

\noindent
\\
\textbf{\large 5.} \textbf{\large Mode radial envelope plays an essential role.} 
\\

The model of \cref{pzdot-2} is used to understand the nonlinear evolution of the particle trajectories (\cref{fig:tracking-pzeta}). According to \cref{pzdot-2,dfdt}, $\dot{\delta f} \propto \dot{\psi}$. Thus, the perturbed distribution function (\cref{fig:deltf-pzeta}) of the resonant particles is similar to the particle trajectories (\cref{fig:tracking-pzeta}).
As shown in \cref{fig:tracking-pzeta}b,
particle trajectories are determined by the oscillations along the $\pmb {\delta\psi}$ and $\pmb {\delta \zeta}$ directions plus the motion in the negative and positive $\pmb {\delta\zeta}$ directions for phases between $\alpha=(0,\pi)$ and $\alpha=(\pi,2\pi)$, respectively.
As the mode grows,
the structure of particle trajectories gradually starts twisting and generating the $\dot{\delta\psi}=0$ turning points.
This persistent twisting eventually leads to the generation of the phase space island (\cref{fig:tracking-pzeta} a\textrightarrow d),
when the perturbed density (\cref{fig:deltf-pzeta}d) and the electric field (\cref{fig:phi-snap}d) both reach their maxima.
Due to the asymmetric radial dependence of the toroidal precessional drift, the trajectories in the phase space island are asymmetric.
At this moment, almost all the resonant particles driving the instabilities are trapped in the phase space island.

Due to the asymmetric radial dependence of the toroidal precessional drift along with the effect from the oscillating electric field, the island continues its twisting (\cref{fig:tracking-pzeta} e\textrightarrow f),
which causes the outer part of the phase island and the inner part of the neighboring island to twist and form a new island.
During this stage, the clumps and holes gradually line up vertically in the radial direction so that the mode structure is destroyed
and the density perturbation becomes small at any toroidal angle.
This leads to the perturbed density (\cref{fig:deltf-pzeta} e\textrightarrow f) and electric field (\cref{fig:phi-snap} e\textrightarrow f) decaying gradually.
Since $\dot{\delta\zeta}$ inside the $q=2$ rational surface is larger than that outside (\cref{fig:tracking-pzeta} e\textrightarrow f),
the weight evolution inside is faster than that outside (\cref{fig:deltf-pzeta} e\textrightarrow f).
Therefore, when the mode amplitude decreases, the perturbed distribution of the outer part decreases more slowly than the inner part, which leads to the mode structure moving outward radially (\cref{fig:phi-snap}e).
When the distribution function in the resonance region outside the rational surface dominates the perturbed distribution function, the precessional frequency $\omega_d$ $(\propto \mu B/r)$ of the resonant particles decreases, corresponding to the reduction of the mode frequency.  
This stage is characterized by a weakening BAE with the broken mode structure and the down-chirping frequency.
\linebreak
\noindent
\\
\textbf{\large 6.} \textbf{\large Particle dynamics determines mode evolution.} 
\\

When the BAE amplitude reaches its minimal level,
the drive from the energetic particles is restored and begins to excite the BAE instability again,
and the process of (f\textrightarrow i) proceeds as opposite to the process of (d\textrightarrow f).
This leads to the repetitive process of mode excitation and suppression accompanied by the down-chirping in the frequency (f\textrightarrow i, i\textrightarrow k).
This repetitive process corresponds to the folding process of the phase space islands, which form
the stretched-nested-folding-layer structure (\cref{fig:tracking-pzeta} f\textrightarrow i, i\textrightarrow k; \cref{fig:deltf-pzeta} f\textrightarrow i, i\textrightarrow k). 
During this folding process, the resonance region of the particle phase space and the mode structure extend radially beyond the rational surface. Comparing with the 1D Landau damping model\cite{o1971nonlinear}, the radial envelope plays an important role. Therefore, our paradigm reveals the intrinsically 2D dynamics in the collisionless nonlinear Landau damping.

Our new paradigm of 2D wave-particle interaction can be extended to other Alfv\'en eigenmodes in toroidal geometry with radial variations of mode amplitude and radially asymmetric particle dynamics. 
For example, the reversed magnetic field has the shear reversed point, so the evolution of the radial dependence of the toroidal precessional drift is different from current analysis, which can be the explanation for the RSAE's \cite{berk2001theoretical, pinches2004role} chirping behavior.
And within our theoretical framework, considering additional physics such as the sideband effect, ballooning structure, thermal particle kinetic effect and so on, we can foresee abundant physics. Finally, our paradigm suggestions, for the first time, that the perpendicular dependence of the the wave-particle interaction plays an essential role for the nonlinear dynamics,
which could have implications beyond the fusion application.

\noindent
\\
\textbf{\large Acknowledgments} 
\\

Authors gratefully acknowledge useful discussions with Dr.\ W.\ Chen and Dr. H. S. Zhang.
This work was supported by the \ITERCN\ under Grant Nos.\
2018YFE0304100,
2018YEF0311300
and 2017YFE0301300;
the \NSFC\ under Grant Nos.\
12025508,
11835016,
11875067,
11675256,
11675257,
\SPRP\ under Grant No.\ XDB16010300;
\KRPFS\ under Grant No.\ QYZDJ-SSW-SYS016;
and SciDAC ISEP.
Resources used in this research were provided by the National Supercomputer Center in Tianjin (NSCC-TJ), the Oak Ridge Leadership Computing Facility at Oak Ridge National Laboratory (OLCF) and the National Energy Research Scientific Computing Center (NERSC).


\begin{thebibliography}{26}
\bibitem{ikeda2007progress} K. Ikeda, Nucl. Fusion \textbf{47}, E01 (2007).
\bibitem{strait1993stability} E. J. Strait, W. W. Heidbrink, A. D. Turnbull, M. S. Chu, and H. H. Duong, Nucl. Fusion \textbf{33}, 1849 (1993).
\bibitem{podesta2011non} M. Podest{\`{a}}, R. E. Bell, N. A. Crocker, E. D. Fredrickson, N. N. Gorelenkov, W. W. Heidbrink, S. Kubota, B. P. LeBlanc, and H. Yuh, Nucl. Fusion \textbf{51}, 063035 (2011).
\bibitem{garcia2011fast} M. Garcia-Munoz, I. G. J. Classen, B. Geiger, W. W. Heidbrink, M. A. Van Zeeland, S. Äkäslompolo, R. Bilato, V. Bobkov, M. Brambilla, G. D. Conway, S. da Gra{\c{c}}a, V. Igochine, Ph. Lauber, N. Luhmann, M. Maraschek, F. Meo, H. Park, M. Schneller, G. Tardini, and the ASDEX Upgrade Team, Nucl. Fusion \textbf{51}, 103013 (2011).
\bibitem{chen2016physics} L. Chen and F. Zonca, Rev. Mod. Phys. \textbf{88}, 015008 (2016).
\bibitem{pinches2004spectroscopic} S. D. Pinches, H. L. Berk, M. P. Gryaznevich, S. E. Sharapov, and JET-EFDA Contributors, Plasma Phys. Control. Fusion \textbf{46}, S47 (2004).
\bibitem{gryaznevich2006perturbative} M. P. Gryanzevich and S. E. Sharapov, Nucl. Fusion \textbf{46}, S942 (2006).
\bibitem{heidbrink2006weak} W. W. Heidbrink, E. Ruskov, E. D. Fredrickson, N. Gorelenkov, S. S. Medley, H. L. Berk, and R. W. Harvey, Plasma Phys. Control. Fusion \textbf{48}, 1347 (2006).
\bibitem{berk1996nonlinear} H. L. Berk, B. N. Breizman, and M. Pekker, Phys. Rev. Lett. \textbf{76}, 1256 (1996).
\bibitem{lilley2009destabilizing} M. K. Lilley, B. N. Breizman, and S. E. Sharapov, Phys. Rev. Lett. \textbf{102}, 195003 (2009).
\bibitem{candy1999nonlinear} J. Candy, H. L. Berk, B. N. Breizman, and F. Porcelli, Phys. Plasmas \textbf{6}, 1822 (1999).
\bibitem{vann2007} R. G. L. Vann, H.L. Berk, and A.R. Soto-Chavez, Phys. Rev. Lett. \textbf{99}, 025003 (2007).
\bibitem{hezaveh2020long} H. Hezaveh, Z. S. Qu, B. N. Breizman, and M. J. Hole, Nucl. Fusion \textbf{60}, 056014 (2020).
\bibitem{zhang2012nonlinear} H. S. Zhang, Z. Lin, and I. Holod, Phys. Rev. Lett. \textbf{109}, 025001 (2012).
\bibitem{wang2012nonlinear} X. Wang, S. Briguglio, L. Chen, C. Di Troia, G. Fogaccia, G. Vlad, and F. Zonca, Phys. Rev. E \textbf{86}, 045401(R) (2012).
\bibitem{zonca2015nonlinear} F. Zonca, L. Chen, S. Briguglio, G. Fogaccia, G. Vlad, and X. Wang, New J. Phys. \textbf{17}, 013052 (2015).
\bibitem{lin1998turbulent} Z. Lin, T. S. Hahm, W. W. Lee, W. M. Tang, and R. B. White, Science \textbf{281}, 1835 (1998).
\bibitem{jcheng2016} J. Cheng, W. Zhang, Z. Lin, I. Holod, D. Li, Y. Chen, and J. Cao, Phys. Plasmas \textbf{23}, 052504 (2016).
\bibitem{white2013theory} R. B. White, The theory of toroidally confined plasmas (World Scientific Pusblishing Co Inc. 2013).
\bibitem{brizard2007foundations} A. J. Brizard and T. S. Hahm, Rev. Mod. Phys. \textbf{79}, 421 (2007).
\bibitem{pliu2017a} P. Liu, W. Zhang, C. Dong, J. Lin, Z. Lin, and J. Cao, Nucl. Fusion \textbf{57}, 126011 (2017).
\bibitem{pliu2017b} P. Liu, W. Zhang, C. Dong, J. Lin, Z. Lin, J. Cao, and D. Li, Phys. Plasmas \textbf{24}, 112114 (2017).
\bibitem{holod2009electromagnetic} I. Holod, W. L. Zhang, Y. Xiao, and Z. Lin, Phys. Plasmas \textbf{16}, 122307 (2009).
\bibitem{greenslade1993all} T. B. Greenslade Jr., Phys. Teach. \textbf{31}, 364 (1993).
\bibitem{o1971nonlinear} T. M. O'Neil, J. H. Winfrey, and J. H. Malmberg, Phys. Fluids \textbf{14}, 1204 (1971).
\bibitem{berk2001theoretical} H. L. Berk, D. N. Borba, B. N. Breizman, S. D. Pinches, and S. E. Sharapov, Phys. Rev. Lett. \textbf{87}, 185002 (2001).
\bibitem{pinches2004role} S. D. Pinches, H. L. Berk, D. N. Borba, B. N. Breizman, S. Brguglio, A. Fasoli, G. Fogaccia, M. P. Gryanzevich, V. Kiptily, M. J. Mantsinen, S. E. Sharapov, D. Testa, R. G. L. Vann, G. Vlad, F. Zonca, and JET-EFDA Contributors, Plasma Phys. Control. Fusion \textbf{46}, B187 (2004).

\end{thebibliography}
\end{document}